%% file: sms_IntEl_v3_resub.tex
\documentclass[aps,pra,floatfix,superscriptaddress,notitlepage]{revtex4-1}
\usepackage[utf8]{inputenc}
\usepackage[OT1]{fontenc}
\usepackage{amsmath}
\usepackage{amsfonts}
\usepackage{amssymb}
\usepackage{bm}
\usepackage{graphicx}
\usepackage[left=2cm,right=2cm,top=2cm,bottom=2cm]{geometry}
\usepackage{longtable}
\allowdisplaybreaks

\usepackage{dcolumn}
\usepackage{siunitx}
\usepackage{multirow}

\usepackage{color}
\definecolor{BLUE}{rgb}{0.0,0.0,1.0}

\usepackage{soul}
\soulregister\cite7
\soulregister\ref7

\newcommand{\veps}{\varepsilon}
\newcommand{\balpha}{\bm{\alpha}}
\newcommand{\bnabla}{\bm{\nabla}}

\newcommand{\bA}{{\rm{\bf{A}}}}
\newcommand{\bD}{\bm{D}}
\newcommand{\br}{\bm{r}}
\newcommand{\bx}{\bm{x}}
\newcommand{\by}{\bm{y}}

\newcommand{\bp}{\bm{p}}

\newcommand{\be}{\begin{eqnarray}}
\newcommand{\ee}{\end{eqnarray}}

\newcommand{\non}{\nonumber \\[2mm]}

\newcommand{\psum}{\sideset{}{'}\sum}
\newcommand{\matr}[3]{\langle #1 | #2 | #3 \rangle}

\begin{document}

\title{QED theory of the specific mass shift in atoms}

%

\author{A. V. Malyshev}
\affiliation{Department of Physics, St.~Petersburg State University, Universitetskaya 7/9, 199034 St.~Petersburg, Russia  
\looseness=-1}

\author{I. S. Anisimova}
\affiliation{Department of Physics, St.~Petersburg State University, Universitetskaya 7/9, 199034 St.~Petersburg, Russia  
\looseness=-1}

\author{D. V. Mironova}
\affiliation{Department of Physics, St.~Petersburg State University, Universitetskaya 7/9, 199034 St.~Petersburg, Russia  
\looseness=-1}
\affiliation{St.~Petersburg Electrotechnical University, Prof. Popov 5, 197376 St.~Petersburg, Russia
\looseness=-1}

\author{V. M. Shabaev}
\affiliation{Department of Physics, St.~Petersburg State University, Universitetskaya 7/9, 199034 St.~Petersburg, Russia  
\looseness=-1}

\author{G. Plunien}
\affiliation{Institut f\"ur Theoretische Physik, Technische Universit\"at Dresden, Mommsenstra{\ss}e 13, D-01062 Dresden, Germany
\looseness=-1}


\begin{abstract}

The formalism of quantum electrodynamics for treating the interelectronic-interaction correction of first order in $1/Z$ to the two-electron part of the nuclear recoil effect on binding energies in atoms and ions is developed. The nonperturbative (in $\alpha Z$) calculations of the corresponding contribution to the energies of the $1s^2$ state in He-like and the $1s^2 2s$ and $1s^2 2p_{1/2}$ states in Li-like ions are performed in the range $Z=5-100$. The behavior of the two-electron part of the nuclear recoil effect beyond the lowest-order relativistic approximation as a function of $Z$ is studied.

\end{abstract}


\maketitle


\section{Introduction \label{sec:0}}

Within the Breit approximation, the nuclear recoil effect on binding energies in atoms and ions can be treated by employing the mass shift (MS) Hamiltonian~\cite{Shabaev:1985:588, Shabaev:1988:69, Palmer:1987:5987} which is given by [the relativistic units ($\hbar=1$, $c=1$) are used throughout the paper]
\begin{align}
\label{eq:H_M}
H_{M} = \frac{1}{2M} \sum_{i,j} \left\{ 
        \bp_i \cdot \bp_j  
      - \frac{\alpha Z}{r_i} \left[ \balpha_i + \frac{(\balpha_i \cdot \br_i) \br_i }{r_i^2} \right] \cdot \bp_j 
                             \right\} \, ,
\end{align}
where the indices $i$ and $j$ enumerate the electrons, $\balpha$ are the Dirac matrices, $\br$ is the position vector, $r=|\br|$, $\bp$ is the momentum operator, $\alpha$ is the fine-structure constant, $Z$ and $M$ are the nuclear charge number and nuclear mass, respectively. The first term in the curly braces in Eq.~(\ref{eq:H_M}) represents the nonrelativistic recoil operator whereas the second term corresponds to the lowest-order relativistic correction. The Hamiltonian~(\ref{eq:H_M}) can be written as a sum of its one- and two-electron parts 
\begin{align}
\label{eq:NMS+SMS}
H_{M}= H_{\rm NMS} + H_{\rm SMS} \, , 
\end{align} 
where 
\begin{align}
\label{eq:NMS}
H_{\rm NMS} = \frac{1}{2M} \sum_{i} \left\{ 
        \bp_i^2
      - \frac{\alpha Z}{r_i} \left[ \balpha_i + \frac{(\balpha_i \cdot \br_i) \br_i }{r_i^2} \right] \cdot \bp_i 
                             \right\}  
\end{align} 
is the normal mass shift (NMS) operator, and
\begin{align}
\label{eq:SMS}
H_{\rm SMS} = \frac{1}{2M} \sum_{i\neq j} \left\{ 
        \bp_i \cdot \bp_j  
      - \frac{\alpha Z}{r_i} \left[ \balpha_i + \frac{(\balpha_i \cdot \br_i) \br_i }{r_i^2} \right] \cdot \bp_j 
                             \right\}
\end{align} 
is the specific mass shift (SMS) operator. The terms ``NMS'' and ``SMS'' sometimes refer only to the nonrelativistic parts of the operators~(\ref{eq:NMS}) and (\ref{eq:SMS}). In this case, the corresponding relativistic corrections given by the second terms in curly braces in Eqs.~(\ref{eq:NMS}) and (\ref{eq:SMS}) are labeled with ``RNMS'' and ``RSMS'', respectively, which denote the relativistic NMS and SMS operators. In the following, we will not separate these contributions employing, e.g., the term SMS for the whole operator~(\ref{eq:SMS}).

The MS operator~(\ref{eq:H_M}) is widely employed nowadays in relativistic calculations of the atomic electronic structure and, especially, isotope shifts (see, e.g., Refs.~\cite{Tupitsyn:2003:022511, SoriaOrts:2006:103002, Korol:2007:022103, Kozhedub:2010:042513, Gaidamauskas:2011:175003, Zubova:2014:062512, Naze:2014:1197, Zubova:2016:052502, Filippin:2017:042502, Gamrath:2018:38, Tupitsyn:2018:022517, Ekman:2019:433} and references therein). The Hamiltonian~$H_M$ allows one to take into account the nuclear recoil corrections within the $(m/M)(\alpha Z)^4mc^2$ approximation. The fully relativistic theory of the nuclear recoil effect to all orders in $\alpha Z$ can be formulated only in the framework of quantum electrodynamics (QED)~\cite{Shabaev:1985:588, Shabaev:1988:69, Shabaev:1998:59, Pachucki:1995:1854, Yelkhovsky:recoil, Adkins:2007:042508}. For the point-nucleus case, the calculations of the QED recoil contributions to the binding energies of few-electron ions to all orders in $\alpha Z$ were performed in Refs.~\cite{Artemyev:1995:1884, Artemyev:1995:5201, Adkins:2007:042508}. The finite nuclear size correction for these terms was partly taken into account for the $1s$ and $2s$ states of H-like ions in Refs.~\cite{Shabaev:1998:4235, Shabaev:1999:493}. We note that the rigorous treatment of the latter correction is currently accessible only within the lowest-order relativistic approximation~\cite{Grotch:1969:350, Borie:1982:67, Aleksandrov:2015:144004}. The most accurate to-date evaluation of the QED recoil effect for all of the $n=1$ and $n=2$ states of He-like ions was made in Ref.~\cite{Malyshev:2018:085001}. The results of the calculations for Be- and B-like ions were presented, e.g., in Refs.~\cite{SoriaOrts:2006:103002, Zubova:2016:052502}. It is worth noting that for high-$Z$ systems the QED recoil corrections can be of comparable magnitude to the values obtained within the Breit approximation. For instance, the total nuclear recoil correction for the ground-state energy of H-like uranium constitutes $0.46$~eV~\cite{Shabaev:1998:4235}, and only about a half of this result comes from the MS operator~(\ref{eq:H_M}). 

All the previous calculations of the nuclear recoil contributions to all orders in $\alpha Z$, see Refs.~\cite{Artemyev:1995:1884, Artemyev:1995:5201, Shabaev:1998:4235, Shabaev:1999:493, SoriaOrts:2006:103002, Adkins:2007:042508, Zubova:2016:052502, Malyshev:2018:085001} and references therein, were limited by the independent-electron approximation, i.e., the interelectronic-interaction effects were treated only to zeroth order in $1/Z$. The present study aims at further development of the QED theory of the nuclear recoil effect in atoms. Namely, we derive the formalism for the QED evaluation of the interelectronic-interaction correction of first order in $1/Z$ to the two-electron part of the nuclear recoil effect on binding energies. The calculations of the two-electron contribution are generally more complicated than the evaluation of the one-electron part, which can be taken into account within the nonrelativistic approximation simply by replacing the electron mass~$m$ with the reduced one, $m_r=mM/(m+M)$. In some sense, the contribution under consideration provides the QED correction for the SMS operator~(\ref{eq:SMS}). In spite of the scaling factor of $1/Z$, this term may significantly contribute to some specific differences of the energies or isotope shifts, see, e.g., the related discussion of the nuclear recoil effect on the bound-state $g$ factor in Ref.~\cite{Malyshev:2017:765}. Moreover, these calculations allow one to better understand the limits of the applicability of the MS Hamiltonian~(\ref{eq:H_M}) for systems where the correlation effects are of great importance, e.g., for many-electron atoms and ions. For instance, to date we have some discrepancies between high-precision measurements and preliminary theoretical predictions for the isotope shifts of the fine-structure splittings in singly ionized calcium (${\rm Ca}^+$)~\cite{Shi:2016:2} and argon (${\rm Ar}^+$)~\cite{Botsi:ICPEAC, Botsi:MS_thesis}. We can assume that a more rigorous QED treatment is necessary in order to resolve these discrepancies. To illustrate all these points, the formalism developed is employed to calculate the two-electron part of the nuclear recoil effect on the energies of the $1s^2$ state in He-like ions and the $1s^2 2s$ and $1s^2 2p_{1/2}$ states in Li-like ions in the wide range $Z=5-100$. The behavior of the nontrivial QED correction to the SMS with increasing $Z$ is analyzed. We note that for the $S$ states, $1s^2$ and $1s^2 2s$, the SMS vanishes to zeroth order in $1/Z$. Therefore, the correction of interest represents the leading two-electron contribution to the nuclear recoil effect for these states.

The paper is organized as follows. In Sec.~\ref{sec:1} we remind the basic ideas of the QED theory of the nuclear recoil effect to zeroth order in $1/Z$. In Sec.~\ref{sec:2} we consider the formulas derived for calculations within the rigorous QED approach of the first-order interelectronic-interaction correction to the two-electron part of the nuclear recoil effect on atomic binding energies. In Sec.~\ref{sec:3} the numerical results are presented and compared with the values obtained within the Breit approximation.



\section{QED theory of the nuclear recoil effect to zeroth~order~in~$1/Z$ \label{sec:1}}

In the present study we start with the QED theory of the nuclear recoil effect in atoms~\cite{Shabaev:1985:588, Shabaev:1988:69} which was generalized in Ref.~\cite{Shabaev:1998:59}. The theory formulated in Ref.~\cite{Shabaev:1998:59} leads to the diagram technique which represents a convenient approach for constructing the QED perturbation series. Within this approach, there is no need to sum infinite sequences of the Feynman diagrams describing the electron-nucleus interaction. This theory will be used in the next section in order to obtain formal expressions for the interelectronic-interaction correction to the two-electron part of the QED recoil effect. However, first, we briefly remind the basic formalism of the theory.

We consider the QED system which in addition to the electron-positron and electromagnetic fields includes also the nucleus. The latter one is assumed to be a nonrelativistic particle with mass $M$ and charge $Z|e|$ ($e<0$ is the electron charge). Since the nuclear recoil effect on energy levels does not depend on the nuclear spin to first order in $m/M$, we consider the nucleus to be spinless. Being an integral of motion, the total momentum of the whole system conserves. Therefore, in the center-of-mass frame the operator of the nuclear momentum can be expressed in terms of the electron-positron-field and electromagnetic-field momenta. Plugging the expression obtained into the Hamiltonian of the whole system, one can derive a field operator~${\rm H}_M$. This operator has to be added to the standard QED Hamiltonian of the electron-positron field interacting with the quantized electromagnetic field and with the classical Coulomb potential of the nucleus, $V_{\rm nucl}$, in order to take into account the nuclear recoil corrections to first order in~$m/M$ and to all orders in~$\alpha Z$. The contributions of first and higher orders in $\alpha$ are beyond the scope of the present study. For this reason, the nontrivial terms involving the electromagnetic-field momentum 
$
{\rm{\bf P}}_f = \int \! d\bx \, [ \mbox{\boldmath$\cal E$}_t(\bx)
                       \times \mbox{\boldmath$\cal H$}  (\bx) ] 
$
contributing to these orders can be discarded in ${\rm H}_M$ actually, see the details in Ref.~\cite{Shabaev:1998:59}. Within this approximation, the operator~${\rm H}_M$ in the Schr\"odinger representation and the Coulomb gauge reads as follows
\begin{align}
\label{eq:dH}
{\rm H}_M &= \frac{1}{2M} \int \! d\bx \, \Psi^\dagger(\bx) (-i\bnabla_{\bx}) \Psi(\bx) 
                        \int \! d\by \, \Psi^\dagger(\by) (-i\bnabla_{\by}) \Psi(\by)    \nonumber \\
         & -\frac{eZ}{M} \int \! d\bx \, \Psi^\dagger(\bx) (-i\bnabla_{\bx}) \Psi(\bx) \, \bA(0) 
           +\frac{e^2Z^2}{2M} \bA(0)^2 \, ,
\end{align}
where $\Psi$ and $\bA$ are the electron-positron and electromagnetic field operators, respectively.

Being interested in the QED theory to all orders in~$\alpha Z$, we employ the Furry picture of QED~\cite{Furry:1951:115}, where the interaction with the classical field of the nucleus is included in the unperturbed Hamiltonian. The perturbation series are constructed by applying the two-times Green function (TTGF) method~\cite{TTGF}. In order to account for the nuclear recoil effect, we take the operator ${\rm H}_M$ in the interaction representation and add it to the interaction part of the Hamiltonian. The Feynman rules for the theory without ${\rm H}_M$ are given, e.g., in Ref.~\cite{TTGF}. The inclusion of the term~${\rm H}_M$ adds several new lines and vertices to the diagram technique, see Ref.~\cite{Shabaev:1998:59} for the details. To introduce the notations employed in the following, we briefly discuss the new elements of the diagram technique by the example of the two-electron contribution.

\begin{figure}
\begin{center}
\includegraphics[width=0.60\textwidth]{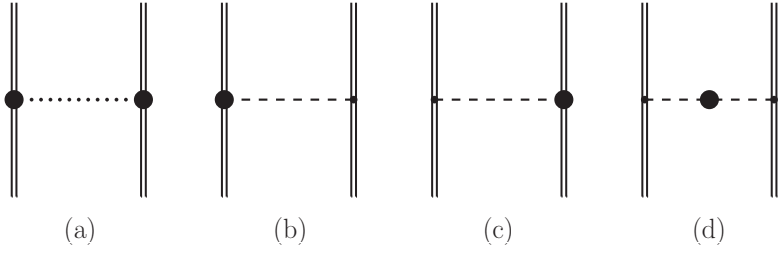}
\caption{\label{fig:recoil_2el}
Two-electron nuclear recoil diagrams to zeroth order in $1/Z$: the Coulomb~(a), one-transverse~(b) and (c), and two-transverse~(d) contributions. See the text and Ref.~\cite{Shabaev:1998:59} for the description of the Feynman rules.}
\end{center}
\end{figure}

To zeroth order in $1/Z$, the two-electron contribution to the nuclear recoil effect on binding energies of a few-electron atom is described by the diagrams shown in Fig.~\ref{fig:recoil_2el}. As usual for bound-state QED, the double line denotes the electron propagator in the classical field of the nucleus. The vertex with a small black dot is the standard vertex of QED. The additional vertices with the bold dots come from the term~${\rm H}_M$ and include the momentum operator $\bp=-i\bnabla$. In accordance with Ref.~\cite{Shabaev:1998:59}, the dotted line ended by two bold dots in Fig.~\ref{fig:recoil_2el}(a) designates the ``Coulomb recoil'' interaction. The dashed lines attached to a bold dot on one side in Figs.~\ref{fig:recoil_2el}(b) and \ref{fig:recoil_2el}(c) denote the ``one-transverse recoil'' interaction, because these lines contain the transverse part of the photon propagator taken in the Coulomb gauge
\begin{align}
\label{eq:D_lk}
D_{lk}(\omega,\br) = 
-\frac{1}{4\pi} \left[ 
\frac{\exp\left( i \sqrt{\omega^2 + i0} \, r \right)}{ r } \delta_{lk}
+
\nabla_l \nabla_k
\frac{\exp\left( i \sqrt{\omega^2 + i0} \, r \right) - 1}{\omega^2 r}
\right] \, ,
\end{align}
where $r=|\br|$ and the branch of the square root is fixed with the condition $\Im\left(\sqrt{\omega^2 + i0}\right)>0$. Finally, the dashed line with a bold dot on it [in Fig.~\ref{fig:recoil_2el}(d)] contains the product of two photon propagators~(\ref{eq:D_lk}) and, for this reason, corresponds to the ``two-transverse recoil'' interaction. We note that the employed separation of the terms as well as the terminology itself result from operating in the Coulomb gauge which is the most convenient one for dealing with the nuclear recoil effect, see, e.g., Refs.~\cite{Shabaev:1985:588, Shabaev:1988:69, Yelkhovsky:recoil}.


Applying the TTGF method, one can easily derive the formulas for the two-electron contribution. For simplicity, we consider a two-electron ion described by the one-determinant unperturbed wave function
\be
\label{eq:u_2el} 
u_{\rm 2el}=\frac{1}{\sqrt{2}} \sum_P (-1)^P 
\psi_{Pa} (\br_1) \psi_{Pb} (\br_2)  \, ,
\ee
where $\psi_n$ are the solutions of the one-electron Dirac equation with the potential of the nucleus included
\begin{equation}
\label{eq:DirEq}
\left[-i \balpha \cdot \nabla + \beta m +V_{\rm nucl}(r)\right] \psi_n ( \br ) = \veps_n \psi_n ( \br ) \, ,
\end{equation}
$P$ is the permutation operator, and $(-1)^P$ is the sign of the permutation. A more general case of an $N$-electron atom described by a many-determinant wave function can be treated in the same manner. According to Ref.~\cite{TTGF}, the first-order correction to the energy of a single level is given by
\begin{align}
\label{eq:dE1}
\Delta E^{(1)}  
=
\frac{1}{2\pi i} \oint_\Gamma \! dE \, \Delta E \Delta g^{(1)}_{uu}(E) \, ,
\end{align}
where $\Delta g^{(1)}_{uu}$ is the Fourier transform of the relevant first-order contribution to two-time Green's function projected on the unperturbed state~(\ref{eq:u_2el}), $\Delta E = E-E_{u}^{(0)}$, and $E_{u}^{(0)}$ is the unperturbed energy. The contour $\Gamma$ oriented counterclockwise has to surround the point $E_{u}^{(0)}$. The derivation of the formulas for the two-electron part of the nuclear recoil effect to zeroth order in $1/Z$ is similar to that of the one-photon exchange correction, see, e.g., Ref.~\cite{Shabaev:1993:4703}. Employing the TTGF method, we obtain
\begin{align}
\label{eq:dE_c}
\Delta E^{(1)}_{\rm c} = 
\frac{1}{M} \sum_P (-1)^P 
\matr{Pa}{p_k}{a} 
\matr{Pb}{p_k}{b} 
\end{align}
for the Coulomb contribution in Fig.~\ref{fig:recoil_2el}(a),
\begin{align}
\label{eq:dE_tr1}
\Delta E^{(1)}_{\rm tr1} = 
-\frac{1}{M} \sum_P (-1)^P 
\left[
\matr{Pa}{p_k}{a} 
\matr{Pb}{D_k(\Delta)}{b} 
+ 
\matr{Pa}{D_k(\Delta)}{a} 
\matr{Pb}{p_k}{b} 
\right]
\end{align}
for the one-transverse-photon contribution in Figs.~\ref{fig:recoil_2el}(b) and \ref{fig:recoil_2el}(c), and
\begin{align}
\label{eq:dE_tr2}
\Delta E^{(1)}_{\rm tr2} = 
\frac{1}{M} \sum_P (-1)^P 
\matr{Pa}{D_k(\Delta)}{a} 
\matr{Pb}{D_k(\Delta)}{b} 
\end{align}
for the two-transverse-photon contribution in Fig.~\ref{fig:recoil_2el}(d). In Eqs.~(\ref{eq:dE_c})-(\ref{eq:dE_tr2}), the summation over the repeated indices is implied  (this convention is held for the subsequent expressions as well), $\Delta=\veps_{Pa}-\veps_a$, and
\begin{align}
\label{eq:D}
D_k(\omega) = -4\pi \alpha Z \alpha_l D_{lk} (\omega) \, ,
\end{align}
where $\alpha_l$ $(l=1,2,3)$ are the Dirac matrices. The total two-electron contribution to the nuclear recoil effect to zeroth order in $1/Z$ is given by the sum of Eqs.~(\ref{eq:dE_c})-(\ref{eq:dE_tr2}),
\begin{align}
\label{eq:dE_rec}
\Delta E^{(1)}_{\rm rec,2el} = \Delta E^{(1)}_{\rm c} + \Delta E^{(1)}_{\rm tr1} + \Delta E^{(1)}_{\rm tr2} \, .
\end{align}
Taking into account Eq.~(\ref{eq:D_lk}), the zero-energy-transfer limit $\omega\to0$ of Eq.~(\ref{eq:D}) reads as
\begin{align}
\label{eq:D0}
D_k(0) = \frac{\alpha Z}{2r} 
\left[
\alpha_k + \frac{(\alpha_i r_i) r_k}{r^2}
\right] \, .
\end{align}
By discarding the two-transverse-photon contribution and considering the limit $\omega\to0$ in the one-transverse-photon term in Eq.~(\ref{eq:dE_rec}), one derives the effective two-electron operator which describes the nuclear recoil effect within the Breit approximation. Obviously, this procedure leads to the SMS operator given in Eq.~(\ref{eq:SMS}). The Coulomb contribution~(\ref{eq:dE_c}) corresponds to the nonrelativistic two-electron recoil operator while its low-order relativistic correction arises from the one-transverse-photon contribution. 


\section{Interelectronic-interaction correction to the two-electron part of the nuclear recoil effect \label{sec:2}}

According to Ref.~\cite{TTGF}, the second-order correction for energy of a single level is given by
\begin{align}
\label{eq:dE2}
\Delta E^{(2)}  
= &
\frac{1}{2\pi i} \oint_\Gamma \! dE \, \Delta E \Delta g^{(2)}_{uu}(E) \,   
-      
\left(
\frac{1}{2\pi i} \oint_\Gamma \! dE \, \Delta E \Delta g^{(1)}_{uu}(E) 
\right)
\left(
\frac{1}{2\pi i} \oint_\Gamma \! dE \, \Delta g^{(1)}_{uu}(E) 
\right) \, ,
\end{align}
where the contour $\Gamma$ surrounds the pole of the level under consideration~$E^{(0)}_u$ and keeps outside all the other singularities of Green's function~$\Delta g^{(2)}_{uu}$. The second term in Eq.~(\ref{eq:dE2}), which we refer to as the disconnected one, usually can be fully canceled by separating the corresponding contributions in the most nontrivial first term. The procedure of the analytical cancellation of the disconnected contribution demands rather tedious manipulations and depends on the total number of electrons~$N$. In this work, we consider the cases of heliumlike ($N=2$) and lithiumlike ($N=3$) ions and present the formulas only for single levels described by one-determinant unperturbed wave functions. The two-electron unperturbed wave function was given in Eq.~(\ref{eq:u_2el}) while in case of $N=3$ the wave function can be written as
\be
\label{eq:u_3el} 
u_{\rm 3el}=\frac{1}{\sqrt{3!}} \sum_P (-1)^P 
\psi_{P1} (\br_1) \psi_{P2} (\br_2) \psi_{P3} (\br_3) \, ,
\ee
where the one-electron states are labeled with the indices $1$, $2$, and $3$. The generalization to the case of a many-determinant wave function is straightforward. Moreover, the derived formalism suits for any atomic systems actually and can be generalized to describe the nuclear recoil effect on energy levels of (quasi-)degenerate states~\cite{TTGF}.


\begin{figure}
\begin{center}
\includegraphics[width=0.60\textwidth]{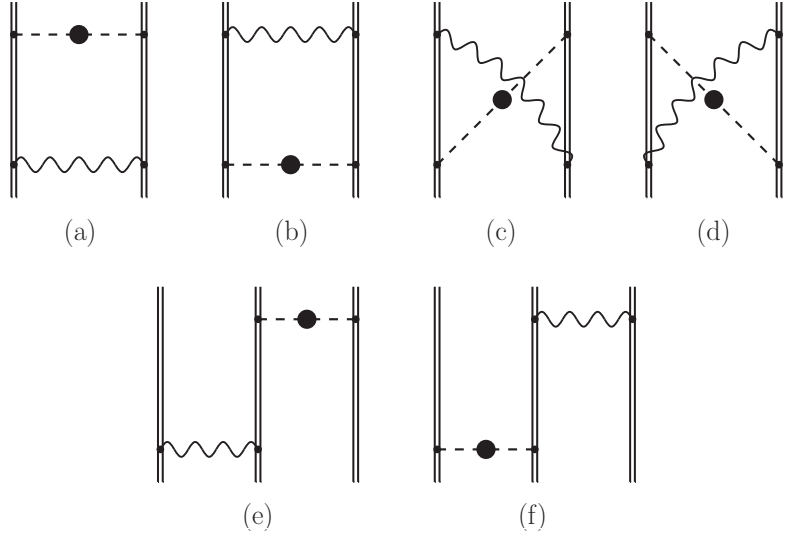}
\caption{\label{fig:recoil_2el_IntEl}
The second-order diagrams describing the interelectronic-interaction correction to the two-electron two-transverse-photon contribution to the nuclear recoil effect. The analogous diagrams with the Coulomb and one-transverse photon recoil interactions have to be taken into account as well. See the text and Ref.~\cite{Shabaev:1998:59} for the description of the diagram technique. 
}
\end{center}
\end{figure}

\begin{figure}
\begin{center}
\includegraphics[width=0.125\textwidth]{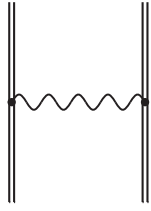}
\caption{\label{fig:1ph}
The one-photon exchange diagram which along with the first-order diagrams in Fig.~\ref{fig:recoil_2el} contributes to the second ``disconnected'' term in Eq.~(\ref{eq:dE2}). 
}
\end{center}
\end{figure}

The example of diagrams describing the interelectronic-interaction correction to the two-electron part of the nuclear recoil effect is shown in Fig.~\ref{fig:recoil_2el_IntEl}. The wavy line denotes the photon propagator here. Other notations are the same as in Fig.~\ref{fig:recoil_2el}. In Fig.~\ref{fig:recoil_2el_IntEl}, only the two-transverse-photon contribution is presented. One should consider also the diagrams with the two-transverse-photon recoil interaction replaced with the Coulomb and one-transverse-photon recoil interactions. As a result, the total number of the second-order diagrams is four times higher actually. We refer to the diagrams in Figs.~\ref{fig:recoil_2el_IntEl}(a) and \ref{fig:recoil_2el_IntEl}(b) as the ladder contribution and to the diagrams in Figs.~\ref{fig:recoil_2el_IntEl}(c) and \ref{fig:recoil_2el_IntEl}(d) as the crossed contribution. For heliumlike ions, only these two-electron diagrams contribute. For lithiumlike ions, the three-electron diagrams in Figs.~\ref{fig:recoil_2el_IntEl}(e) and \ref{fig:recoil_2el_IntEl}(f) come into play as well. The list of diagrams, which have to be accounted for in the disconnected term in Eq.~(\ref{eq:dE2}), includes the first-order diagrams in Fig.~\ref{fig:recoil_2el} and the one-photon-exchange diagram shown in Fig.~\ref{fig:1ph}.


For the subsequent consideration, it is convenient to introduce the following notations
\begin{align}
\label{eq:I}
I(\omega) &= e^2 \alpha_1^\mu \alpha_2^\nu D_{\mu\nu}(\omega) \, ,   \\
\label{eq:R_c}
R_{\rm c} &= \frac{1}{M} \, \bp_1 \cdot \bp_2 \, ,     \\
\label{eq:R_tr1}
R_{\rm tr1}(\omega) &= -\frac{1}{M} \, \big[ \bp_1 \cdot \bD_2(\omega) + \bD_1(\omega) \cdot \bp_2 \big] \, ,  \\
\label{eq:R_tr2}
R_{\rm tr2}(\omega) &= \frac{1}{M} \, \bD_1(\omega) \cdot \bD_2(\omega) \, ,   
\end{align}
where $\alpha^\mu = (1,\balpha)$, $D_{\mu\nu}$ is the photon propagator, and the vector $\bD$ was defined in Eq.~(\ref{eq:D}). We imply also that $I'(\omega)=dI(\omega)/d\omega$ and $R'(\omega)=dR(\omega)/d\omega$, where $R$ means any of the operators~(\ref{eq:R_c})-(\ref{eq:R_tr2}). In the Coulomb gauge employed, Eq.~(\ref{eq:I}) reads as follows
\begin{align}
\label{eq:I_coul}
I(\omega) &= \alpha \, \Bigg[ \,
\frac{1}{r_{12}} - \frac{(\balpha_1\cdot\balpha_2)\exp\left( i \sqrt{\omega^2 + i0} \, r_{12} \right)}{r_{12}}   
+ 
(\balpha_1\cdot\bnabla_1)
(\balpha_2\cdot\bnabla_2) \,
\frac{\exp\left( i \sqrt{\omega^2 + i0} \, r_{12} \right)-1}{\omega^2 r_{12}} 
\, \Bigg] \, .
\end{align}
From Eqs.~(\ref{eq:D}) and (\ref{eq:I_coul}), it is obvious that in the Coulomb gauge the following symmetry properties $I(\omega)=I(-\omega)$ and $R(\omega)=R(-\omega)$ are held. For brevity, we will designate the matrix elements of the operators (\ref{eq:I}) and (\ref{eq:R_c})-(\ref{eq:R_tr2}) as $I_{abcd}(\omega)=\matr{ab}{I(\omega)}{cd}$ and $R_{abcd}(\omega)=\matr{ab}{R(\omega)}{cd}$, respectively. The zero-energy-transfer limit $\omega \to 0$ of Eq.~(\ref{eq:I_coul}) which along with the MS operator (\ref{eq:H_M}) can be employed to evaluate the effects of the interelectronic interaction on the nuclear recoil within the Breit approximation is given by 
\begin{align}
\label{eq:I0}
I &= \alpha \, \Bigg[ \,
\frac{1}{r_{12}} - \frac{(\balpha_1\cdot\balpha_2)}{r_{12}}   
- 
\frac{(\balpha_1\cdot\bnabla_1)(\balpha_2\cdot\bnabla_2)\, r_{12} }{2} 
\, \Bigg] \, .
\end{align}

The derivation of the formal expressions for the interelectronic-interaction correction to the two-electron part of the nuclear recoil effect within the TTGF method is very similar to the derivation of the corresponding formulas for the two-photon exchange contribution which was considered in details in Refs.~\cite{Shabaev:1994:4489, Yerokhin:2001:032109}. We present only the final expressions omitting all the intermediate steps. First, we discuss the contribution of the two-electron diagrams presented in Figs.~\ref{fig:recoil_2el_IntEl}(a)-(d) and the related diagrams with the Coulomb and one-transverse-photon recoil interactions. As noted above, the two-electron diagrams provide the total result in case of heliumlike ions. On the other hand, the three-electron problem with the unperturbed wave function~(\ref{eq:u_3el}) can be decomposed into three two-electron problems of the type~(\ref{eq:u_2el}). Therefore, the two-electron contribution has to be taken into account for all possible electron pairs $(ab)=(12), (13)$,~and~$(23)$ in the three-electron state~$u_{\rm 3el}$. The contribution of the ladder (``lad'') diagrams in Figs.~\ref{fig:recoil_2el_IntEl}(a) and \ref{fig:recoil_2el_IntEl}(b) is divided naturally into irreducible (``irr'') and reducible (``red'') parts. The reducible part covers the terms for which an intermediate-state energy coincides with the energy~$E_u^{(0)}=\veps_a+\veps_b$ of the state under consideration whereas the irreducible part includes the remainder. The irreducible part of the ladder diagrams reads as
\begin{align}
\label{eq:lad_irr}
\Delta E^{(2)}_{\rm lad,irr} =
\sum_P (-1)^P 
\psum_{n_1n_2}
\sum_{\mu_{n_1}\mu_{n_2}} \, 
\frac{i}{2\pi} \int \! d\omega \, 
\left[
\frac{ I_{PaPb\,n_1n_2}(\omega) R_{n_1n_2ab}(\omega-\veps_{Pa}+\veps_a) }
{\big( \veps_{Pa} - \omega - u\veps_{n_1} \big)
\big( \veps_{Pb} + \omega - u\veps_{n_2} \big) } 
+ \{ I \leftrightarrow R \}
\right] \, ,
\end{align}
where  $u=(1-i0)$ provides the proper treatment of the poles in the electron propagator, and the prime on the sum indicates that the intermediate states with $\veps_{n_1}+\veps_{n_2}=\veps_a+\veps_b$ are excluded. As to the reducible part, the condition $\veps_{n_1}+\veps_{n_2}=\veps_a+\veps_b$ generally restricts the summation over $n_1$ and $n_2$ to the terms with $(\veps_{n_1}\veps_{n_2})=(\veps_a\veps_b),(\veps_b\veps_a)$. However, since the matrix elements of the operators $\bp$ and $\bD$ are equal to zero for states which have the same parity, one can conclude that only one of these possibilities contributes. For the same reason, the reducible part of the ladder diagrams does not vanish identically as a whole only if the electrons $a$ and $b$ belong to different electron shells having the opposite parity. The reducible part of the ladder diagram can be expressed as
\begin{align}
\label{eq:lad_red}
\Delta E^{(2)}_{\rm lad,red} &= 
\frac{1}{2}
\sum_P (-1)^P
\sum_{\mu_{\tilde{a}}\mu_{\tilde{b}}} \,
\left(\frac{-i}{2\pi}\right) \int \! d\omega \, 
\frac{1}{(\omega+i0)^2}  \non 
& \times
\Big[\,
I_{PaPb\,\tilde{b}\tilde{a}}(\omega+\veps_{Pa}-\veps_{b}) R_{\tilde{b}\tilde{a}ab}(\omega+\veps_a-\veps_{b})
+
I_{PaPb\,\tilde{b}\tilde{a}}(\omega+\veps_{Pb}-\veps_{a}) R_{\tilde{b}\tilde{a}ab}(\omega+\veps_b-\veps_{a})    \non 
& \,\,\,\,
+
R_{ab\,\tilde{b}\tilde{a}}(\omega+\veps_a-\veps_b) I_{\tilde{b}\tilde{a}PaPb}(\omega+\veps_{Pa}-\veps_{b})
+
R_{ab\,\tilde{b}\tilde{a}}(\omega+\veps_b-\veps_a) I_{\tilde{b}\tilde{a}PaPb}(\omega+\veps_{Pb}-\veps_{a})
\,\Big] \, ,
\end{align}
where it is assumed that $\veps_{\tilde{a}}=\veps_a$ and $\veps_{\tilde{b}}=\veps_b$.  
Finally, the contribution of the crossed (``cr'') diagrams in Figs.~\ref{fig:recoil_2el_IntEl}(c) and \ref{fig:recoil_2el_IntEl}(d) is given by
\begin{align}
\label{eq:cr}
\Delta E^{(2)}_{\rm cr} =
\sum_P (-1)^P 
\sum_{n_1n_2} \sum_{\mu_{n_1}\mu_{n_2}} \,
\frac{i}{2\pi} \int \! d\omega \,
\left[ 
\frac{I_{Pa\,n_2n_1b}(\omega) R_{n_1Pb\,an_2}(\omega-\veps_{Pa}+\veps_a)}
{\big( \veps_{Pa} - \omega - u\veps_{n_1} \big)   
\big( \veps_{b} - \omega - u\veps_{n_2} \big)}
+ \{ I \leftrightarrow R \}
\right] \, .
\end{align}

Now, we consider the contribution of the three-electron diagrams in Figs.~\ref{fig:recoil_2el_IntEl}(e) and \ref{fig:recoil_2el_IntEl}(f). As in case of the ladder diagrams, one can divide the three-electron contribution into the irreducible and reducible parts. The irreducible contribution of the three-electron diagrams reads as
\begin{align}
\label{eq:3el_irr}
\Delta E^{(2)}_{\rm 3el,irr} =
\sum_{PQ} (-1)^{P+Q} \psum_n \,
\left[
\frac{I_{P2P3nQ3}(\Delta_{P3Q3}) R_{P1nQ1Q2}(\Delta_{Q1P1})}
{\veps_{Q1}+\veps_{Q2}-\veps_{P1}-\veps_n} 
+ \{ I \leftrightarrow R \}
\right] \, ,
\end{align}
where the prime on the sum indicates that the terms with vanishing denominator have to be omitted in the summation. The contribution of the reducible part of the three-electron diagrams in Figs.~\ref{fig:recoil_2el_IntEl}(e) and \ref{fig:recoil_2el_IntEl}(f) can be expressed as
\begin{align}
\label{eq:3el_red}
\Delta E^{(2)}_{\rm 3el,red} &=
\frac{1}{2}\,\sum_{PQ} (-1)^{P+Q} \!\!\!
\sum_{\veps_n=\veps_{Q1}+\veps_{Q2}-\veps_{P1}}   
\Big[ \,
  I'_{P2P3nQ3}(\Delta_{P3Q3}) R_{P1nQ1Q2}(\Delta_{Q1P1})     \non 
& \qquad\qquad\qquad\qquad\qquad\qquad\quad\,\,\,
+ I_{P2P3nQ3}(\Delta_{P3Q3}) R'_{P1nQ1Q2}(\Delta_{Q1P1})  
+ \{ I \leftrightarrow R \}
\, \Big] \, ,
\end{align}

To summarize, in case of a single level in heliumlike ion the QED interelectronic-interaction correction of first order in $1/Z$ to the two-electron part of the nuclear recoil effect is given by the sum of Eqs.~(\ref{eq:lad_irr})-(\ref{eq:cr}). For lithiumlike ions, in order to take into account the corresponding correction one has to calculate Eqs.~(\ref{eq:lad_irr})-(\ref{eq:cr}) for all possible pairs of electrons present in the unperturbed three-electron state and then add the contribution of Eqs.~(\ref{eq:3el_irr}) and (\ref{eq:3el_red}). The calculations have to be performed for all the operators~(\ref{eq:R_c})-(\ref{eq:R_tr2}),
\begin{align}
\label{eq:dE_rec2}
\Delta E^{(2)}_{\rm rec,2el} = \Delta E^{(2)}_{\rm c} + \Delta E^{(2)}_{\rm tr1} + \Delta E^{(2)}_{\rm tr2} \, .
\end{align}

Finally, we note that the formalism presented in this section reproduces the expressions for the interelectronic-interaction correction to the SMS within the Breit approximation if one neglects the energy dependence in the operators $\bD(\omega)$ and $I(\omega)$ in Eqs.~(\ref{eq:D}) and (\ref{eq:I_coul}), respectively, and introduces projectors on the positive-energy part of the spectrum. As previously, the Coulomb gauge is implied for the interelectronic-interaction operator $I(\omega)$, so that within the zero-energy-transfer limit one comes to the operator $I$ in Eq.~(\ref{eq:I0}). On these assumptions, all the reducible contributions vanish since $I'(0)=0$ and $R'(0)=0$, and the $\omega$ integrations in the two-electron terms can be carried out analytically employing Cauchy's residue theorem. The contribution of the crossed diagram vanishes because all the zeros of the denominators in Eq.~(\ref{eq:cr}) lie in the upper half-plane and, therefore, the integration contour can be closed in the lower half-plane avoiding the singularities. Therefore, the irreducible part of the ladder contribution yields the total two-electron correction within the Breit approximation:
\begin{align}
\label{eq:2el_Breit}
\Delta E_{\rm 2el,Breit}^{(2)} = 
\sum_P (-1)^P \psum_{n_1n_2} \sum_{\mu_{n_1}\mu_{n_2}}
\left[ 
\frac{I_{PaPb\,n_1n_2}(0) R_{n_1n_2ab}(0)}{\veps_a+\veps_b-\veps_{n_1}-\veps_{n_2}}
+ \{ I \leftrightarrow R \}
\right] \, ,
\end{align}
where the summation over $n_1$ and $n_2$ is restricted by the conditions $\veps_{n_1}>0$, $\veps_{n_2}>0$, and $\veps_{n_1}+\veps_{n_2}\neq \veps_a+\veps_b$. The three-electron contribution within the Breit approximation is readily obtained from Eq.~(\ref{eq:3el_irr}) by discarding the negative-energy part of the spectrum $\veps_n<0$ and replacing $\Delta_{P3Q3}$ and $\Delta_{Q1P1}$ with zeros. The contribution of the two-transverse-photon operator~(\ref{eq:R_tr2}) has to be omitted within this approximation.


\section{Numerical results and discussion \label{sec:3}}

In the present section, the formalism derived in Secs.~\ref{sec:1} and \ref{sec:2} is applied to the all-order (in $\alpha Z$) evaluation of the two-electron contribution to the nuclear recoil effect on the binding energies of the $1s^2$ state in heliumlike ions and the $1s^2 2s$ and $1s^2 2p_{1/2}$ states in lithiumlike ions. In Ref.~\cite{Shabaev:1998:59}, it was shown that the nuclear size correction to the nuclear recoil effect can be partially taken into account by replacing the pure Coulomb potential $V_{\rm nucl}=-\alpha Z/r$ with the potential of an extended nucleus. Following this prescription, we employ the Fermi model to describe the nuclear charge distribution for all ions except for the ones with $Z=5$ and $Z=10$. For the latter nuclei, the homogeneously-charged-sphere model is used instead. The nuclear charge radii are taken from Refs.~\cite{Angeli:2013:69, Yerokhin:2015:033103}. The summation over intermediate electron states is performed employing the finite basis sets constructed from the B-splines~\cite{Johnson:1988:307, Sapirstein:1996:5213} within the dual kinetic balance approach~\cite{splines:DKB}. 

For states under consideration, to zeroth order in $1/Z$ the two-electron recoil contribution does not vanish only for the state $1s^2 2p_{1/2}$. The results of our calculations expressed in terms of the dimensionless function $A(\alpha Z)$,
\begin{equation}
\label{eq:A_alphaZ}
\Delta E^{(1)}_{\rm rec,2el} = \frac{m}{M}(\alpha Z)^2 A(\alpha Z) \, mc^2 \, ,
\end{equation}
are given in Table~\ref{tab:0:p:bind_1s1s2p1}. We stress that the index ``$(1)$'' in the left part of Eq.~(\ref{eq:A_alphaZ}) designates that the corresponding energy shift is obtained as the first-order perturbation within the TTGF method. For each $Z$, the values evaluated according to Eqs.~(\ref{eq:dE_c})-(\ref{eq:dE_tr2}) are shown in the first line. The results obtained within the lowest-order relativistic approximation employing the SMS operator~$H_{\rm SMS}$ are displayed in the second lines. The functions $A_{\rm c}$, $A_{\rm tr1}$, and $A_{\rm tr2}$ correspond to the terms $\Delta E^{(1)}_{\rm c}$, $\Delta E^{(1)}_{\rm tr1}$, and $\Delta E^{(1)}_{\rm tr2}$, respectively. One can see that to zeroth order in $1/Z$ the Coulomb contribution~$A_{\rm c}$ has the same value within the both approaches. The deviation of the one-transverse-photon term is determined by the frequency-dependent correction in the operator $\bD(\omega)$ in Eq.~(\ref{eq:D}). The two-transverse-photon contribution is absent in the Breit approximation. From Table~\ref{tab:0:p:bind_1s1s2p1}, it is seen that the terms of the higher orders in $\alpha Z$ can significantly alter the total values, especially, for high-$Z$ ions, where the contribution of the nonrelativistic part of the SMS operator (\ref{eq:SMS}) is canceled considerably by the contribution due to the low-order relativistic correction for it, see, e.g., the relevant discussion in Ref.~\cite{Tupitsyn:2003:022511}. For the point-nucleus case, the corresponding correction was considered previously in Ref.~\cite{Artemyev:1995:1884}. We note that in Ref.~\cite{Artemyev:1995:1884} the two-electron contribution for the $1s^2 2p_{1/2}$ was presented in terms of the dimensionless function $Q(\alpha Z)$ which differs from the function $A(\alpha Z)$ by the factor of $-3^8/2^9$, see Eq.~(74) in Ref.~\cite{Artemyev:1995:1884}. For comparison, the point-nucleus results from Ref.~\cite{Artemyev:1995:1884} expressed in terms of the function~$A(\alpha Z)$ are given in the last column of Table~\ref{tab:0:p:bind_1s1s2p1}.

The interelectronic-interaction correction of first order in $1/Z$ to the two-electron part of the nuclear recoil effect is conveniently represented via the dimensionless function $B(\alpha Z)$ defined by
\begin{equation}
\label{eq:B_alphaZ}
\Delta E^{(2)}_{\rm rec,2el} = \frac{m}{M}\frac{(\alpha Z)^2}{Z} B(\alpha Z) \, mc^2 \, ,
\end{equation}
The results of the calculations for the $1s^2$, $1s^2 2s$, and $1s^2 2p_{1/2}$ states expressed in terms of the function $B(\alpha Z)$ are presented in Tables~\ref{tab:1:bind_1s1s}, \ref{tab:1:bind_1s1s2s}, and \ref{tab:1:bind_1s1s2p1}, respectively. As in Table~\ref{tab:0:p:bind_1s1s2p1}, for each $Z$ the results of the QED calculations to all orders in $\alpha Z$ as well as the values obtained employing the SMS operator~$H_{\rm SMS}$ are given. The functions $B_{\rm c}$, $B_{\rm tr1}$, and $B_{\rm tr2}$ correspond to the contributions of the Coulomb~(\ref{eq:R_c}), the one-transverse-photon~(\ref{eq:R_tr1}), and the two-transverse-photon~(\ref{eq:R_tr2}) operators, respectively. The uncertainties given in the tables correspond only to errors of the numerical calculations. They were estimated by increasing the size of the employed basis set and also by studying how the integrations over the energy parameter $\omega$ in Eq.~(\ref{eq:lad_irr}) and the other related contributions converge. When the uncertainty is not specified, all the digits presented should be correct. Except for the heaviest ions with $Z \geqslant  92$, the uncertainties due to varying the nuclear charge distribution model as well as the nuclear charge radii are below the number of digits shown. For the heaviest ions, this varying may alter the last digit. In addition, we should stress once more that the calculations with the wave functions evaluated for the extended nucleus correspond to a partial treatment of the nuclear size corrections to the recoil effect. The uncertainty due to this approximation can be estimated in accordance with the prescription given, e.g., in Ref.~\cite{Malyshev:2018:085001}.

As noted at the end of the previous section, the calculation formulas which are valid within the lowest-order relativistic approximation can be obtained from the general QED expressions if we neglect the energy dependence of the transverse part of the photon propagator in the Coulomb gauge in Eq.~(\ref{eq:D_lk}), restrict the consideration to the positive-energy part of the Dirac spectrum, and omit the two-transverse-photon contribution. As an independent crosscheck, we evaluated the two-electron part of the nuclear recoil effect in the Breit approximation employing the numerical code for the QED calculations and compared the results obtained with the direct application of the SMS operator~(\ref{eq:SMS}). The two calculations were found to be in agreement with each other. 

\begin{figure}
\begin{center}
\includegraphics[width=0.77\textwidth]{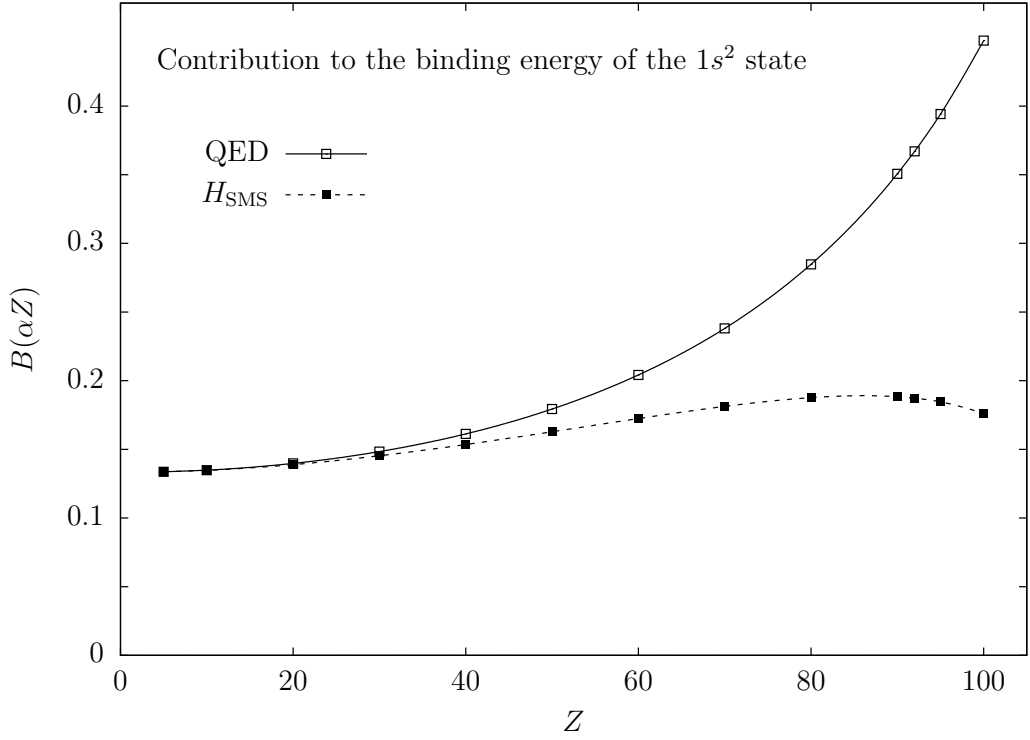}
\caption{\label{fig:1:bind_1s1s}
The first-order in $1/Z$ interelectronic-interaction correction to the two-electron part of the nuclear recoil effect on the binding energy of the $1s^2$ state expressed in terms of the dimensionless function $B(\alpha Z)$ defined by Eq.~(\ref{eq:B_alphaZ}) The solid line represents the results of the QED calculations to all orders in $\alpha Z$ whereas the dashed line stands for the calculations based on the specific mass shift (SMS) operator given by Eq.~(\ref{eq:SMS}). 
}
\end{center}
\end{figure}

\begin{figure}
\begin{center}
\includegraphics[width=0.77\textwidth]{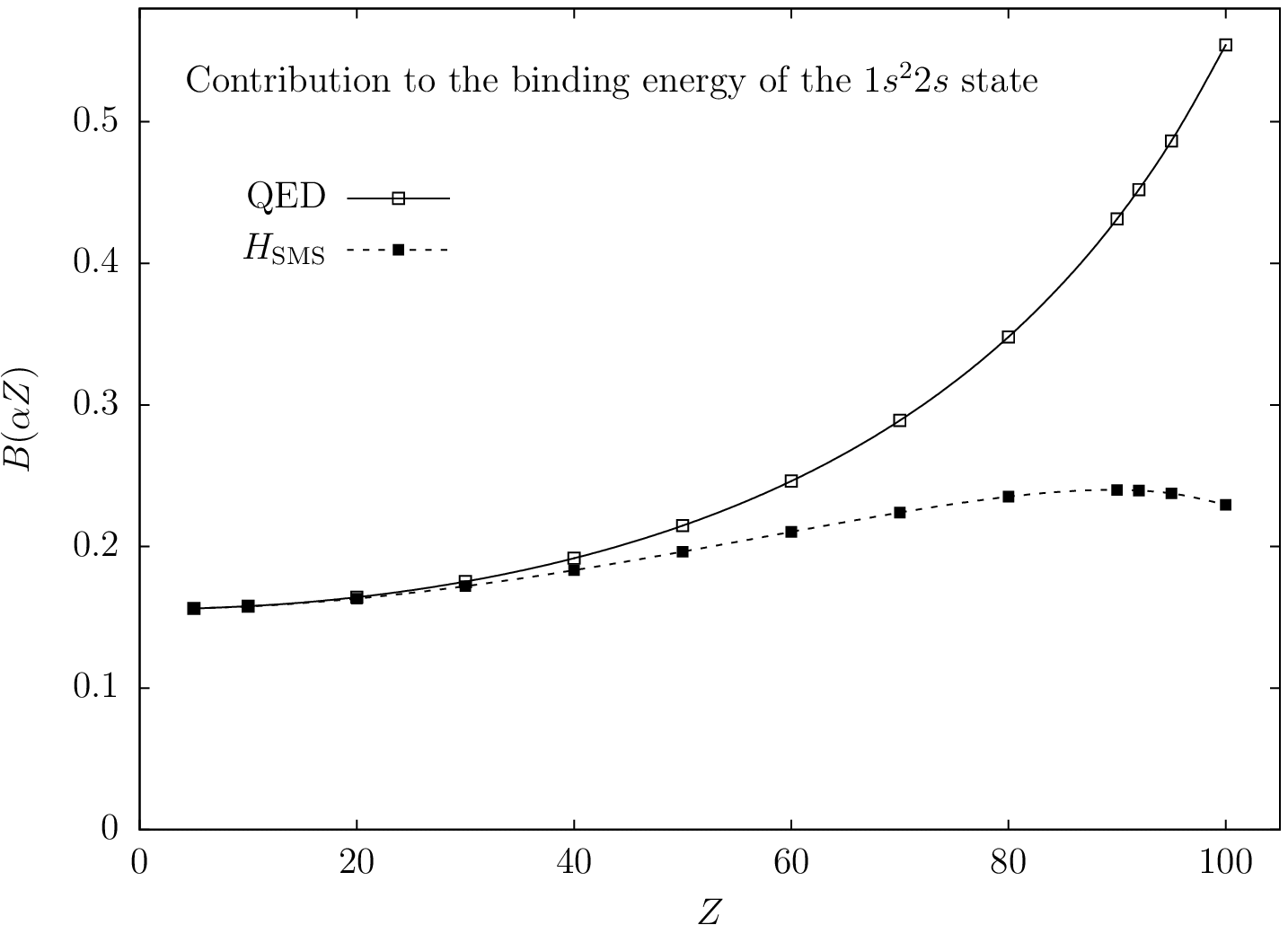}
\caption{\label{fig:1:bind_1s1s2s}
The first-order in $1/Z$ interelectronic-interaction correction to the two-electron part of the nuclear recoil effect on the binding energy of the $1s^2 2s$ state expressed in terms of the dimensionless function $B(\alpha Z)$ defined by Eq.~(\ref{eq:B_alphaZ}). Notations are the same as in Fig.~\ref{fig:1:bind_1s1s}. 
}
\end{center}
\end{figure}

\begin{figure}
\begin{center}
\includegraphics[width=0.77\textwidth]{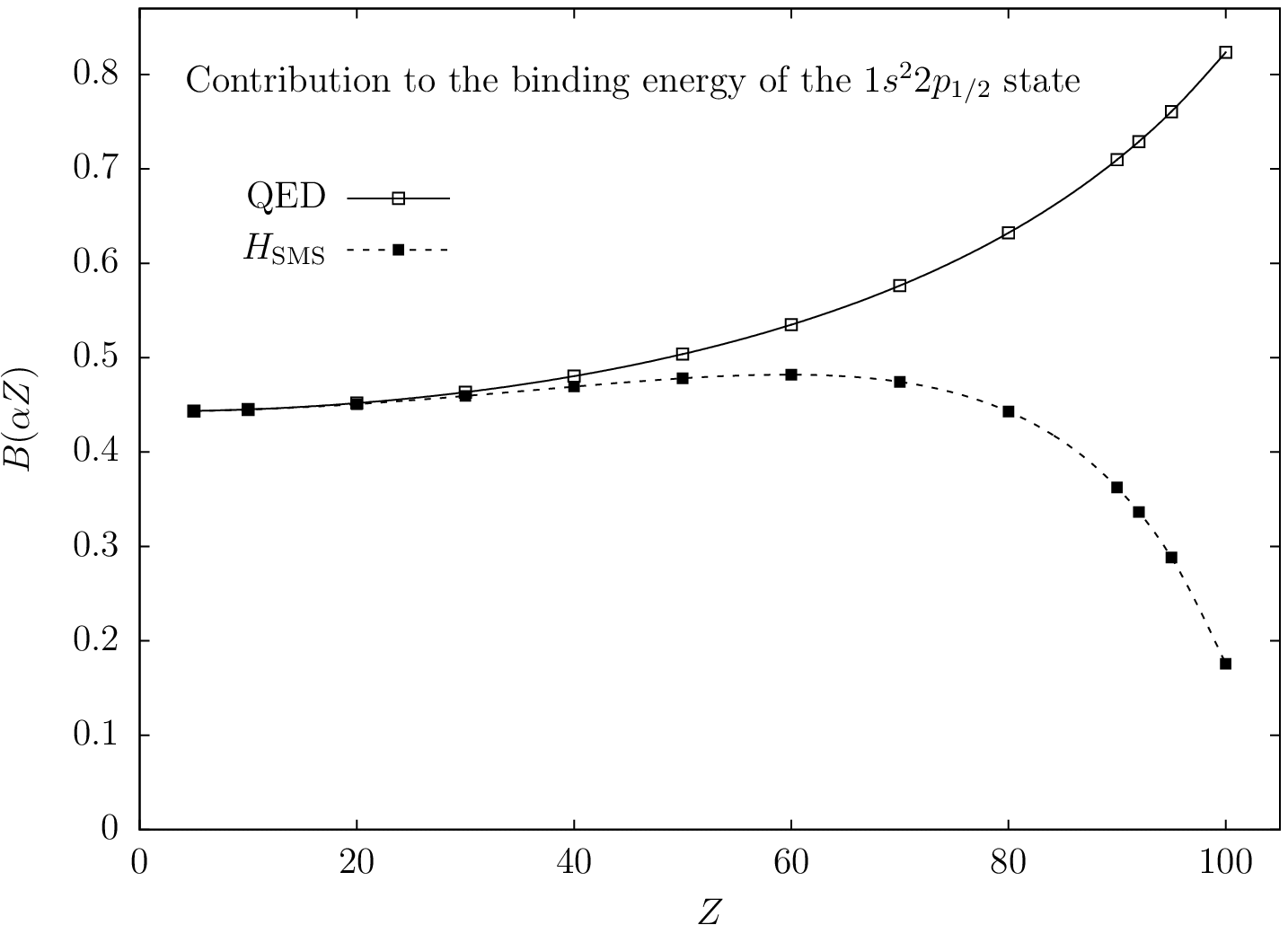}
\caption{\label{fig:1:bind_1s1s2p1}
The first-order in $1/Z$ interelectronic-interaction correction to the two-electron part of the nuclear recoil effect on the binding energy of the $1s^2 2p_{1/2}$ state expressed in terms of the dimensionless function $B(\alpha Z)$ defined by Eq.~(\ref{eq:B_alphaZ}). Notations are the same as in Fig.~\ref{fig:1:bind_1s1s}. 
}
\end{center}
\end{figure}

From Tables~\ref{tab:1:bind_1s1s}-\ref{tab:1:bind_1s1s2p1}, one can note that, compared to the independent-electron approximation, the Coulomb contribution acquires the correction to the Breit-approximation result due to the higher orders in $\alpha Z$. The alteration of the one-transverse-photon contribution is also more pronounced than it takes place to zeroth order in~$1/Z$, since the corresponding correction is not limited to the simple inclusion of the frequency-dependent correction. In addition, the two-transverse-photon contribution increases rapidly with increasing $Z$. As a result, the total QED values may drastically differ from the approximate ones evaluated to lowest orders in $\alpha Z$ employing the operator~$H_{\rm SMS}$. In order to illustrate the behavior of the interelectronic-interaction correction to the two-electron part of the nuclear recoil effect, we plot the total contributions to the binding energies of the states under consideration in Figs.~\ref{fig:1:bind_1s1s}-\ref{fig:1:bind_1s1s2p1}. The data given in the last columns of Tables~\ref{tab:1:bind_1s1s}-\ref{tab:1:bind_1s1s2p1} are presented. The results obtained employing the SMS operator (\ref{eq:SMS}) are shown with dashed lines. The values calculated by means of \textit{ab initio} approach derived in the previous section are displayed with solid lines. It is worth noting that for the $1s^2 2p_{1/2}$ state the interelectronic-interaction correction to the two-electron recoil within the Breit approximation tends to zero as it was found for the leading in $1/Z$ contribution. From Fig.~\ref{fig:1:bind_1s1s2p1}, one can see that taking into account of the effects of higher orders in $\alpha Z$ changes the situation. Finally, we should note also that by combining the data presented in Tables~\ref{tab:1:bind_1s1s}-\ref{tab:1:bind_1s1s2p1} one can readily obtain the interelectronic-interaction correction to the two-electron part of the nuclear recoil effect on the ionization potentials of the $1s^2 2s$ and $1s^2 2p_{1/2}$ states as well as the $2p_{1/2}-2s$ transition energy in lithiumlike ions.

\begin{figure}
\begin{center}
\includegraphics[width=0.77\textwidth]{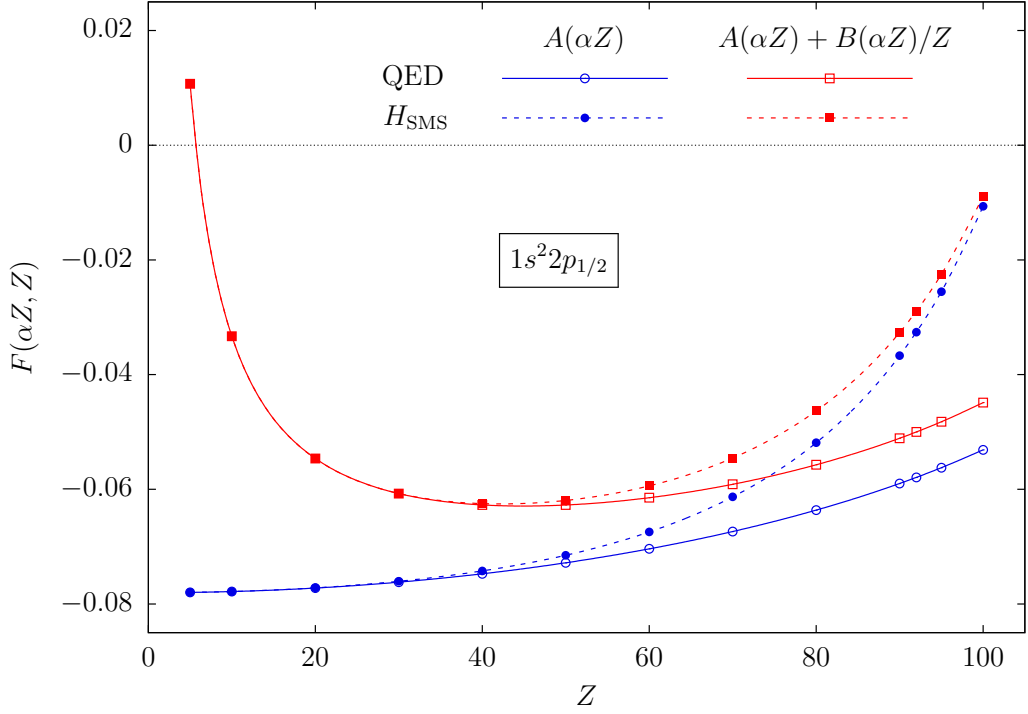}
\caption{\label{fig:01:bind_1s1s2p1} 
The two-electron part of the nuclear recoil effect on the binding energy of the $1s^2 2p_{1/2}$ state expressed in terms of the dimensionless function $F(\alpha Z,Z)$ defined by Eqs.~(\ref{eq:F_alphaZ}) and (\ref{eq:F_alphaZ_sum}). The solid lines represent the results of the QED calculations to all orders in $\alpha Z$ while the dashed lines stand for the calculations based on the specific mass shift (SMS) operator given by Eq.~(\ref{eq:SMS}). The contributions of zeroth order in $1/Z$, $F_0(\alpha Z) = A(\alpha Z)$, and the sums of zeroth and first orders in $1/Z$, $F_{01}(\alpha Z,Z) = A(\alpha Z) + B(\alpha Z)/Z$, are shown with blue (circles) and red (squares) lines, respectively. 
}
\end{center}
\end{figure}

The total two-electron nuclear recoil contribution to the energy shift can be expressed as
\begin{equation}
\label{eq:F_alphaZ}
\Delta E_{\rm rec,2el} = \frac{m}{M}(\alpha Z)^2 F(\alpha Z,Z) \, mc^2 \, ,
\end{equation}
where, in accordance with the definitions given in Eqs.~(\ref{eq:A_alphaZ}) and (\ref{eq:B_alphaZ}), one obtains
\begin{equation}
\label{eq:F_alphaZ_sum}
F(\alpha Z,Z) = A(\alpha Z) + \frac{1}{Z}B(\alpha Z) + \ldots \, , 
\end{equation}
and an elipsis in Eq.~(\ref{eq:F_alphaZ_sum}) corresponds to the terms of the second and higher orders in $1/Z$. As noted above, for the $S$ states, $1s^2$ and $1s^2 2s$, the $1/Z$ perturbation theory starts from the first-order correction~$B(\alpha Z)$, and the contribution of interest represents the leading two-electron term. For the $1s^2 2p_{1/2}$ state, it is not the case. Therefore, in Table~\ref{tab:01:bind_1s1s2p1} we compare the zeroth- and first-order contributions to the corresponding function $F(\alpha Z,Z)$. The term $A(\alpha Z)$ is taken from the penultimate column in Table~\ref{tab:0:p:bind_1s1s2p1} while the function $B(\alpha Z)$ is from the last column in Table~\ref{tab:1:bind_1s1s2p1}. For illustrative purposes, the data given in Table~\ref{tab:01:bind_1s1s2p1} are plotted also in Fig.~\ref{fig:01:bind_1s1s2p1}. As in Figs.~\ref{fig:1:bind_1s1s}-\ref{fig:1:bind_1s1s2p1}, the dashed lines correspond to the calculations with the SMS operator~(\ref{eq:SMS}), and the solid lines represent the QED results. The zeroth-order contributions to the function $F(\alpha Z,Z)$ are indicated with the blue lines with circles on them. The next-to-leading approximations to the function $F(\alpha Z,Z)$, given by the sums of zeroth and first orders in $1/Z$, are shown with the red lines with squares on them. Naturally, for low-$Z$ ions the $1/Z$ perturbation theory may converge slowly. From Fig.~\ref{fig:01:bind_1s1s2p1}, it is seen that the interelectronic-interaction correction to the SMS is comparable in magnitude with the leading contribution. For this reason, our calculations taken alone do not pretend to provide the best possible theoretical predictions for the two-electron part of the nuclear recoil effect for low-$Z$ ions. If needed, the results obtained for these systems can be further improved by considering within the Breit approximation the second- and higher-order contributions to Eq.~(\ref{eq:F_alphaZ}) by means of, e.g., the configuration interaction~\cite{Tupitsyn:2003:022511} or the recursive perturbation theory~\cite{Glazov:2017:46} methods. In the present work, we pursue the aim to study the influence of the nontrivial QED effects on the two-electron recoil contribution. In this regard, one can see from Table~\ref{tab:01:bind_1s1s2p1} and Fig.~\ref{fig:01:bind_1s1s2p1} that taking into account of the terms of higher orders in $\alpha Z$ considerably changes the behavior of the function $F(\alpha Z,Z)$ as a function of $Z$. The calculations based on the SMS operator~$H_{\rm SMS}$ lead to a underestimation of the two-electron contribution for high-$Z$ ions. Moreover, the dashed lines in Fig.~\ref{fig:01:bind_1s1s2p1} lie much closer to each other than the solid ones for high-$Z$ ions. This designates once again that the nontrivial QED contribution of first order in $1/Z$ represents the significant effect. 

\begin{figure}
\begin{center}
\includegraphics[width=0.77\textwidth]{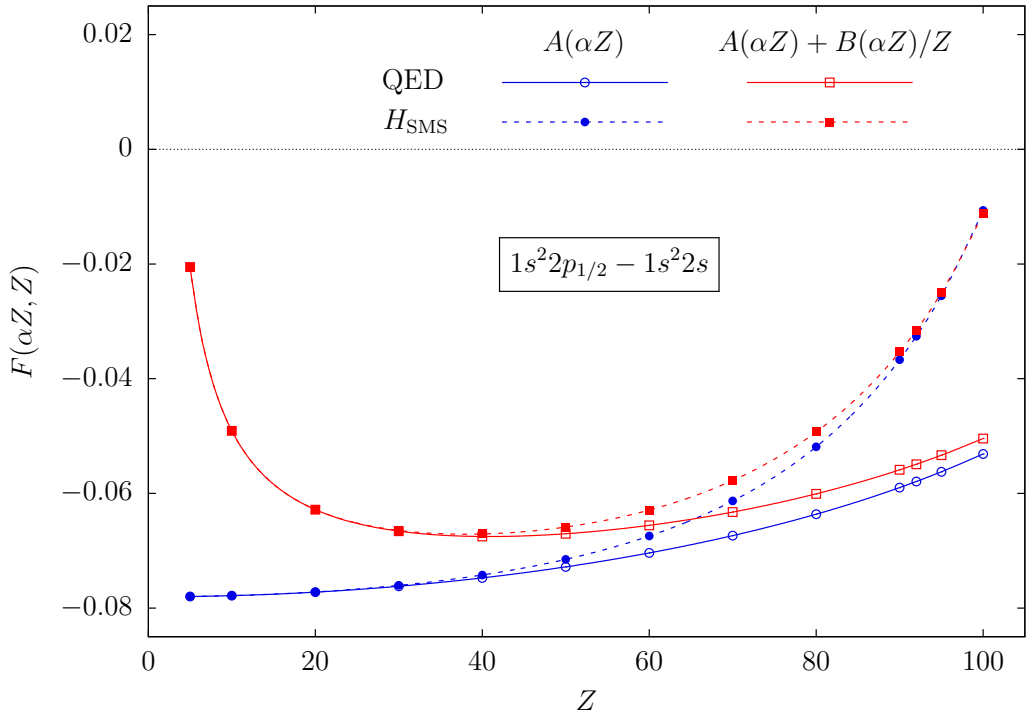}
\caption{\label{fig:01:tr_2p1_2s} 
The two-electron part of the nuclear recoil effect on the $2p_{1/2}-2s$ transition energy in Li-like ions expressed in terms of the dimensionless function $F(\alpha Z,Z)$ defined by Eqs.~(\ref{eq:F_alphaZ}) and (\ref{eq:F_alphaZ_sum}). Notations are the same as in Fig.~\ref{fig:01:bind_1s1s2p1}.
}
\end{center}
\end{figure}

Finally, we consider the two-electron part of the nuclear recoil effect on the $2p_{1/2}-2s$ transition energy in lithiumlike ions. For the point-nucleus case, the one-electron contribution arising from the NMS operator~(\ref{eq:NMS}) can be evaluated analytically to zeroth order in $1/Z$~\cite{Shabaev:1985:588}:
\begin{equation}
\label{eq:HMS_1el}
\Delta E_{\rm rec,1el}^{\rm(p)} = \frac{m^2-\veps^2}{2M} \, , 
\end{equation}
where $\veps$ is the Dirac energy. Since $\veps_{2s}=\veps_{2p_{1/2}}$ for the pure Coulomb potential $V_{\rm nucl} = -\alpha Z/ r$, the one-electron contribution within the Breit approximation vanishes in this limit. Therefore, the total mass shift for this transition is determined by the finite-nuclear-size, one-electron QED as well as two-electron recoil effects. In Fig.~\ref{fig:01:tr_2p1_2s}, we plot the two-electron nuclear recoil contribution to the $2p_{1/2}-2s$ transition energy evaluated by means of the $1/Z$ perturbation theory up to the first order. The notations are the same as in Fig.~\ref{fig:01:bind_1s1s2p1} for the binding energy of the $1s^2 2p_{1/2}$ state. Since the two-electron recoil term for the $1s^2 2s$ state is equal to zero within the independent-electron approximation, to zeroth order in $1/Z$ the corresponding contributions to the transition and $1s^2 2p_{1/2}$ state coincide with each other (the blue lines in Figs.~\ref{fig:01:bind_1s1s2p1} and \ref{fig:01:tr_2p1_2s} are the same). The first-order interelectronic-interaction correction can be obtained by taking the difference of the results presented in Tables~\ref{tab:1:bind_1s1s2p1} and \ref{tab:1:bind_1s1s2s}, respectively. From Figs.~\ref{fig:01:bind_1s1s2p1} and \ref{fig:01:tr_2p1_2s}, one can conclude that, in principle, the behavior of the total two-electron nuclear recoil effect with the growth of $Z$ is rather similar in these two cases. Compared to the binding energy of the $1s^2 2p_{1/2}$ state, the nontrivial QED part of the interelectronic-interaction correction is reduced slightly for the $2p_{1/2}-2s$ transition. Nevertheless, it notably contributes. For instance, in Refs.~\cite{Kozhedub:2010:042513, Zubova:2014:062512} the nuclear recoil correction for the $2p_{1/2}-2s$ transition energy was studied. The approach employed there merges the calculations based on the MS operator~(\ref{eq:H_M}) within the Breit approximation to all orders in $1/Z$ with the QED contributions evaluated within the independent-electron approximation~\cite{Artemyev:1995:1884}. The nuclear recoil corrections were presented in terms of the mass shift coefficient $K$ defined according to
\begin{equation}
\label{eq:K}
\Delta E_{\rm rec} = \frac{K}{M} \, .
\end{equation}
In Refs.~\cite{Kozhedub:2010:042513, Zubova:2014:062512}, the mass shift coefficients for the $2p_{1/2}-2s$ transition energy in lithiumlike thorium and uranium were found to be (in units of 1000~GHz~amu) $K^{\rm Th}=-3441(57)$ and $K^{\rm U} =-3734(65)$, respectively. As noted in Ref.~\cite{Zubova:2014:062512}, the uncertainties specified are mainly due to the estimation of the uncalculated QED contributions of first order in $1/Z$. Based on the results obtained in this work for the interelectronic-interaction correction to the two-electron recoil effect which are presented in Tables~\ref{tab:1:bind_1s1s2s} and \ref{tab:1:bind_1s1s2p1}, one can extract the nontrivial QED part of first order in $1/Z$. This two-electron QED correction constitutes (in units of 1000~GHz~amu) $\delta K^{\rm Th}_{\rm QED,2el}=51$ and $\delta K^{\rm U}_{\rm QED,2el}=60$ for thorium and uranium ions, respectively. The theoretical accuracy of the mass shift calculations for the $2p_{1/2}-2s$ transition can be significantly improved, provided the one-electron QED correction of first order in $1/Z$ is calculated. We should stress that, to zeroth order in $1/Z$, the one- and two-electron QED recoil corrections contribute to the total mass shift for the $2p_{1/2}-2s$ transition with the same sign enhancing each other, see Ref.~\cite{Kozhedub:2010:042513}. If this trend persists in first order in $1/Z$, one may expect that the effect of the uncalculated QED contributions is probably underestimated in Ref.~\cite{Kozhedub:2010:042513, Zubova:2014:062512}. 


\section{Summary \label{sec:4}}

To summarize, we have derived the formalism for \textit{ab initio} calculations of the interelectronic-interaction correction to the two-electron part of the nuclear recoil effect on binding energies in atoms and ions to all orders in $\alpha Z$. The technique developed was applied to evaluate the two-electron recoil contributions for the $1s^2$ state in heliumlike ions and the $1s^2 2s$ and $1s^2 2p_{1/2}$ states in lithiumlike ions in the wide range $Z=5-100$. The corresponding contribution to the $2p_{1/2}-2s$ transition energy in lithiumlike ions was investigated as well. The results of the QED calculations to zeroth and first orders in $1/Z$ were compared with their counterparts obtained by employing the specific mass shift operator $H_{\rm SMS}$ given by Eq.~(\ref{eq:SMS}). The behavior of the nontrivial two-electron QED contribution with increasing nuclear charge number $Z$ was discussed. The obtained all-order (in $\alpha Z$) results allow one to estimate in a more rigorous way the accuracy of the calculations based on the mass shift Hamiltonian $H_{M}$ in Eq.~(\ref{eq:H_M}) which describes the nuclear recoil effects only within the $(m/M)(\alpha Z)^4mc^2$ approximation.  

In the future, we plan to extend the QED formalism developed in order to study the interelectronic-interaction correction to the one-electron part of the nuclear recoil effect on binding energies in atoms. In particular, this will allow one to improve the theoretical accuracy of the mass shift calculations in highly charged ions. We note also that the largest contribution to the theoretical uncertainty of the isotope shift of the g factor in lithiumlike calcium is currently determined by the screened QED contributions of first order in $1/Z$ \cite{Shabaev:2017:263001}. In view of the experiments presently implemented at the Max-Planck-Institut f\"ur Kernphysik (MPIK) in Heidelberg~\cite{Sturm:2017:4} and at GSI in Darmstadt~\cite{Lindenfels:2013:023412, Vogel:2019:1800211}, which are aimed at further improvement of the experimental precision of the $g$ factor itself as well as the isotope shifts of the $g$ factor, the QED calculations of the nuclear recoil effect on the $g$ factor of highly charged ions turn out to be urgent. In this connection, the QED theory of the nuclear recoil effect on binding energies developed represents a good starting point for the corresponding theory for the $g$ factor.

Finally, the nonperturbative (in $\alpha Z$) calculations of the nuclear recoil contributions of first order in $\alpha$ for hydrogen and light hydrogenlike ions are also of great interest. The comparison between the nonperturbative numerical approach and the analytical perturbative techniques may provide important data for the remaining higher-order contributions beyond the known $\alpha Z$-expansion terms, see the related discussion about the contribution of the nuclear recoil effect on the Lamb shift to zeroth order in $\alpha$ in Refs.~\cite{Yerokhin:2015:233002, Yerokhin:2016:062514}. 


\section*{Acknowledgments}

We thank Ivan Aleksandrov, Dmitry Glazov, and Ilya Tupitsyn for valuable discussions. The work was supported by RFBR (Grant No.~18-32-00294). A.V.M. and V.M.S. acknowledge the support from the Foundation for the advancement of theoretical physics and mathematics ``BASIS''. A.V.M. and D.V.M also acknowledge the support from the German-Russian Interdisciplinary Science Center (G-RISC) and DAAD Programm Ostpartnerschaften, TU Dresden.


\newpage
\input{table_0_bind_1s1s2p1_with_point.tex}

\newpage
\input{table_1_bind_1s1s.tex}

\newpage
\input{table_1_bind_1s1s2s.tex}

\newpage
\input{table_1_bind_1s1s2p1.tex}

\newpage
\input{table_01_bind_1s1s2p1.tex}

\clearpage

\clearpage


\end{document}

%% file: table_0_bind_1s1s2p1_with_point.tex
{
\renewcommand{\arraystretch}{1.3}
\begin{longtable}{
                  l
                  c
                  S[table-format=-2.7]
                  S[table-format=-2.7]
                  S[table-format=-2.7]
                  S[table-format=-2.7]
                  S[table-format=-2.6]
                 }
 \caption{\label{tab:0:p:bind_1s1s2p1} 
         The two-electron recoil contribution of zeroth order in $1/Z$ to the binding energy of the $1s^22p_{1/2}$ state
         expressed in terms of the dimensionless function $A(\alpha Z)$ defined by Eq.~(\ref{eq:A_alphaZ}). 
         For each $Z$, the first line shows 
         the results of the QED calculations to all orders in $\alpha Z$, whereas the second line displays the values
         obtained within the Breit approximation employing the specific mass shift (SMS) operator given in Eq.~(\ref{eq:SMS}).
         The results by Artemyev \textit{et. al} \cite{Artemyev:1995:1884} for point-nucleus case expressed in terms of $A(\alpha Z)$ are in the last column.
         }\\
\hline
\hline
 \multicolumn{1}{c}{\rule{0pt}{1.2em}$Z$}  &  \multicolumn{1}{c}{Approach}  &  \multicolumn{1}{c}{$A_{\rm c}(\alpha Z)$}   &  
             \multicolumn{1}{c}{$A_{\rm tr1}(\alpha Z)$}   &  \multicolumn{1}{c}{$A_{\rm tr2}(\alpha Z)$}   &
             \multicolumn{1}{c}{$A(\alpha Z)$}             &  \multicolumn{1}{c}{$A^{\rm (p)}(\alpha Z)$ \cite{Artemyev:1995:1884}}   \\
\hline
\endfirsthead
\caption{\it (Continued.)}\\
\hline
\hline
 \multicolumn{1}{c}{\rule{0pt}{1.2em}$Z$}  &  \multicolumn{1}{c}{Approach}  &  \multicolumn{1}{c}{$A_{\rm c}(\alpha Z)$}   &  
             \multicolumn{1}{c}{$A_{\rm tr1}(\alpha Z)$}   &  \multicolumn{1}{c}{$A_{\rm tr2}(\alpha Z)$}   &
             \multicolumn{1}{c}{$A(\alpha Z)$}             &  \multicolumn{1}{c}{$A^{\rm (p.n.)}(\alpha Z)$ \cite{Artemyev:1995:1884}}   \\
\hline
\endhead
\hline
\endfoot
\hline
\endlastfoot
                       
  \multirow{2}{*}{  5}  &  QED            &      -0.078168   &        0.000182   &       -0.000000   &       -0.077986   &       -0.077986    \\ 
                        &  $H_{\rm SMS}$  &      -0.078168   &        0.000182   &    {\text{---}}   &       -0.077986   &                    \\ 

\hline

  \multirow{2}{*}{ 10}  &  QED            &      -0.078565   &        0.000732   &       -0.000002   &       -0.077835   &       -0.077835    \\ 
                        &  $H_{\rm SMS}$  &      -0.078565   &        0.000732   &    {\text{---}}   &       -0.077833   &                    \\ 

\hline

  \multirow{2}{*}{ 20}  &  QED            &      -0.080186   &        0.002989   &       -0.000028   &       -0.077225   &       -0.077225    \\ 
                        &  $H_{\rm SMS}$  &      -0.080186   &        0.002990   &    {\text{---}}   &       -0.077196   &                    \\ 

\hline

  \multirow{2}{*}{ 30}  &  QED            &      -0.083015   &        0.006960   &       -0.000145   &       -0.076199   &       -0.076199    \\ 
                        &  $H_{\rm SMS}$  &      -0.083015   &        0.006969   &    {\text{---}}   &       -0.076046   &                    \\ 

\hline

  \multirow{2}{*}{ 40}  &  QED            &      -0.087267   &        0.013008   &       -0.000482   &       -0.074741   &       -0.074741    \\ 
                        &  $H_{\rm SMS}$  &      -0.087267   &        0.013033   &    {\text{---}}   &       -0.074234   &                    \\ 

\hline

  \multirow{2}{*}{ 50}  &  QED            &      -0.093301   &        0.021735   &       -0.001254   &       -0.072820   &       -0.072819    \\ 
                        &  $H_{\rm SMS}$  &      -0.093301   &        0.021795   &    {\text{---}}   &       -0.071506   &                    \\ 

\hline

  \multirow{2}{*}{ 60}  &  QED            &      -0.101698   &        0.034138   &       -0.002828   &       -0.070388   &       -0.070385    \\ 
                        &  $H_{\rm SMS}$  &      -0.101698   &        0.034256   &    {\text{---}}   &       -0.067442   &                    \\ 

\hline

  \multirow{2}{*}{ 70}  &  QED            &      -0.113418   &        0.051891   &       -0.005840   &       -0.067367   &       -0.067361    \\ 
                        &  $H_{\rm SMS}$  &      -0.113418   &        0.052091   &    {\text{---}}   &       -0.061327   &                    \\ 

\hline

  \multirow{2}{*}{ 80}  &  QED            &      -0.130121   &        0.077941   &       -0.011452   &       -0.063632   &       -0.063623    \\ 
                        &  $H_{\rm SMS}$  &      -0.130121   &        0.078235   &    {\text{---}}   &       -0.051886   &                    \\ 

\hline

  \multirow{2}{*}{ 90}  &  QED            &      -0.154856   &        0.117812   &       -0.021945   &       -0.058988   &       -0.058972    \\ 
                        &  $H_{\rm SMS}$  &      -0.154856   &        0.118162   &    {\text{---}}   &       -0.036694   &                    \\ 

\hline

  \multirow{2}{*}{ 92}  &  QED            &      -0.161216   &        0.128274   &       -0.024984   &       -0.057926   &       -0.057908    \\ 
                        &  $H_{\rm SMS}$  &      -0.161216   &        0.128619   &    {\text{---}}   &       -0.032597   &                    \\ 

\hline

  \multirow{2}{*}{ 95}  &  QED            &      -0.171943   &        0.146083   &       -0.030375   &       -0.056235   &       -0.056214    \\ 
                        &  $H_{\rm SMS}$  &      -0.171943   &        0.146407   &    {\text{---}}   &       -0.025536   &                    \\ 

\hline

  \multirow{2}{*}{100}  &  QED            &      -0.193788   &        0.182924   &       -0.042259   &       -0.053123   &       -0.053097    \\ 
                        &  $H_{\rm SMS}$  &      -0.193788   &        0.183143   &    {\text{---}}   &       -0.010645   &                    \\

\end{longtable}
}

%% file: table_1_bind_1s1s.tex
{
\renewcommand{\arraystretch}{1.3}
\begin{longtable}{
                  l
                  c
                  S[table-format=-1.5(1)]
                  S[table-format=-1.5(1)]
                  S[table-format=-1.5(1)]
                  S[table-format=-1.5(1)]
                 }
 \caption{\label{tab:1:bind_1s1s} 
         The interelectronic-interaction correction of first order in $1/Z$ to the two-electron part
         of the nuclear recoil contribution to the binding energy of the $1s^2$ state
         expressed in terms of the dimensionless function $B(\alpha Z)$ defined by Eq.~(\ref{eq:B_alphaZ}). 
         }\\
\hline
\hline
 \multicolumn{1}{c}{\rule{0pt}{1.2em}$Z$}  &  \multicolumn{1}{c}{Approach}  &  \multicolumn{1}{c}{$B_{\rm c}(\alpha Z)$}   &  
             \multicolumn{1}{c}{$B_{\rm tr1}(\alpha Z)$}   &  \multicolumn{1}{c}{$B_{\rm tr2}(\alpha Z)$}   &
             \multicolumn{1}{c}{$B(\alpha Z)$}   \\
\hline
\endfirsthead
\caption{\it (Continued.)}\\
\hline
\hline
 \multicolumn{1}{c}{\rule{0pt}{1.2em}$Z$}  &  \multicolumn{1}{c}{Approach}  &  \multicolumn{1}{c}{$B_{\rm c}(\alpha Z)$}   &  
             \multicolumn{1}{c}{$B_{\rm tr1}(\alpha Z)$}   &  \multicolumn{1}{c}{$B_{\rm tr2}(\alpha Z)$}   &
             \multicolumn{1}{c}{$B(\alpha Z)$}   \\
\hline
\endhead
\hline
\endfoot
\hline
\endlastfoot
                       
  \multirow{2}{*}{  5}  &  QED            &        0.13393   &        -0.00025   &         0.00000   &         0.13368    \\ 
                        &  $H_{\rm SMS}$  &        0.13394   &        -0.00029   &    {\text{---}}   &         0.13366    \\ 

\hline

  \multirow{2}{*}{ 10}  &  QED            &        0.13578   &        -0.00092   &         0.00000   &         0.13486    \\ 
                        &  $H_{\rm SMS}$  &        0.13589   &        -0.00116   &    {\text{---}}   &         0.13473    \\ 

\hline

  \multirow{2}{*}{ 20}  &  QED            &        0.14297   &        -0.00326   &         0.00007   &         0.13977    \\ 
                        &  $H_{\rm SMS}$  &        0.14381   &        -0.00494   &    {\text{---}}   &         0.13888    \\ 

\hline

  \multirow{2}{*}{ 30}  &  QED            &        0.15481   &        -0.00678   &         0.00034   &         0.14837    \\ 
                        &  $H_{\rm SMS}$  &        0.15748   &        -0.01216   &    {\text{---}}   &         0.14532    \\ 

\hline

  \multirow{2}{*}{ 40}  &  QED            &        0.17169   &        -0.01151   &         0.00107   &         0.16125    \\ 
                        &  $H_{\rm SMS}$  &        0.17774   &        -0.02422   &    {\text{---}}   &         0.15352    \\ 

\hline

  \multirow{2}{*}{ 50}  &  QED            &        0.19453   &        -0.01779   &         0.00262   &         0.17936    \\ 
                        &  $H_{\rm SMS}$  &        0.20603   &        -0.04322   &    {\text{---}}   &         0.16281    \\ 

\hline

  \multirow{2}{*}{ 60}  &  QED            &        0.22495   &        -0.02633   &         0.00556   &         0.20418    \\ 
                        &  $H_{\rm SMS}$  &        0.24477   &        -0.07234   &    {\text{---}}   &         0.17242    \\ 

\hline

  \multirow{2}{*}{ 70}  &  QED            &        0.26574   &        -0.03850   &         0.01081   &         0.23805    \\ 
                        &  $H_{\rm SMS}$  &        0.29790   &        -0.11662   &    {\text{---}}   &         0.18128    \\ 

\hline

  \multirow{2}{*}{ 80}  &  QED            &        0.32176   &        -0.05694   &         0.01994   &         0.28476    \\ 
                        &  $H_{\rm SMS}$  &        0.37224   &        -0.18457   &    {\text{---}}   &         0.18767    \\ 

\hline

  \multirow{2}{*}{ 90}  &  QED            &        0.40176   &        -0.08701   &         0.03591   &         0.35067    \\ 
                        &  $H_{\rm SMS}$  &        0.48001   &        -0.29167   &    {\text{---}}   &         0.18833    \\ 

\hline

  \multirow{2}{*}{ 92}  &  QED            &        0.42192   &        -0.09527   &         0.04039   &         0.36704    \\ 
                        &  $H_{\rm SMS}$  &        0.50733   &        -0.32006   &    {\text{---}}   &         0.18727    \\ 

\hline

  \multirow{2}{*}{ 95}  &  QED            &        0.45562   &        -0.10967   &         0.04821   &         0.39415    \\ 
                        &  $H_{\rm SMS}$  &        0.55314   &        -0.36854   &    {\text{---}}   &         0.18460    \\ 

\hline

  \multirow{2}{*}{100}  &  QED            &        0.52333   &     -0.14077(1)   &         0.06509   &      0.44764(2)    \\ 
                        &  $H_{\rm SMS}$  &        0.64543   &        -0.46918   &    {\text{---}}   &         0.17625    \\

\end{longtable}
}

%% file: table_1_bind_1s1s2s.tex
{
\renewcommand{\arraystretch}{1.3}
\begin{longtable}{
                  l
                  c
                  S[table-format=-1.5(1)]
                  S[table-format=-1.5(1)]
                  S[table-format=-1.5(1)]
                  S[table-format=-1.5(1)]
                 }
 \caption{\label{tab:1:bind_1s1s2s} 
         The interelectronic-interaction correction of first order in $1/Z$ to the two-electron part
         of the nuclear recoil contribution to the binding energy of the $1s^22s$ state
         expressed in terms of the dimensionless function $B(\alpha Z)$ defined by Eq.~(\ref{eq:B_alphaZ}). %
         }\\
\hline
\hline
 \multicolumn{1}{c}{\rule{0pt}{1.2em}$Z$}  &  \multicolumn{1}{c}{Approach}  &  \multicolumn{1}{c}{$B_{\rm c}(\alpha Z)$}   &  
             \multicolumn{1}{c}{$B_{\rm tr1}(\alpha Z)$}   &  \multicolumn{1}{c}{$B_{\rm tr2}(\alpha Z)$}   &
             \multicolumn{1}{c}{$B(\alpha Z)$}   \\
\hline
\endfirsthead
\caption{\it (Continued.)}\\
\hline
\hline
 \multicolumn{1}{c}{\rule{0pt}{1.2em}$Z$}  &  \multicolumn{1}{c}{Approach}  &  \multicolumn{1}{c}{$B_{\rm c}(\alpha Z)$}   &  
             \multicolumn{1}{c}{$B_{\rm tr1}(\alpha Z)$}   &  \multicolumn{1}{c}{$B_{\rm tr2}(\alpha Z)$}   &
             \multicolumn{1}{c}{$B(\alpha Z)$}   \\
\hline
\endhead
\hline
\endfoot
\hline
\endlastfoot
                       
  \multirow{2}{*}{  5}  &  QED            &        0.15655   &        -0.00028   &         0.00000   &         0.15627    \\ 
                        &  $H_{\rm SMS}$  &        0.15657   &        -0.00032   &    {\text{---}}   &         0.15625    \\ 

\hline

  \multirow{2}{*}{ 10}  &  QED            &        0.15885   &        -0.00104   &         0.00000   &         0.15782    \\ 
                        &  $H_{\rm SMS}$  &        0.15899   &        -0.00131   &    {\text{---}}   &         0.15768    \\ 

\hline

  \multirow{2}{*}{ 20}  &  QED            &        0.16782   &        -0.00372   &         0.00008   &         0.16418    \\ 
                        &  $H_{\rm SMS}$  &        0.16882   &        -0.00558   &    {\text{---}}   &         0.16324    \\ 

\hline

  \multirow{2}{*}{ 30}  &  QED            &        0.18264   &        -0.00778   &         0.00038   &         0.17523    \\ 
                        &  $H_{\rm SMS}$  &        0.18581   &        -0.01384   &    {\text{---}}   &         0.17197    \\ 

\hline

  \multirow{2}{*}{ 40}  &  QED            &        0.20386   &        -0.01334   &         0.00118   &         0.19169    \\ 
                        &  $H_{\rm SMS}$  &        0.21107   &        -0.02782   &    {\text{---}}   &         0.18325    \\ 

\hline

  \multirow{2}{*}{ 50}  &  QED            &        0.23268   &        -0.02085   &         0.00290   &         0.21473    \\ 
                        &  $H_{\rm SMS}$  &        0.24647   &        -0.05014   &    {\text{---}}   &         0.19633    \\ 

\hline

  \multirow{2}{*}{ 60}  &  QED            &        0.27124   &        -0.03123   &         0.00618   &         0.24619    \\ 
                        &  $H_{\rm SMS}$  &        0.29514   &        -0.08481   &    {\text{---}}   &         0.21032    \\ 

\hline

  \multirow{2}{*}{ 70}  &  QED            &        0.32319   &        -0.04622   &         0.01206   &         0.28902    \\ 
                        &  $H_{\rm SMS}$  &        0.36221   &        -0.13825   &    {\text{---}}   &         0.22396    \\ 

\hline

  \multirow{2}{*}{ 80}  &  QED            &        0.39489   &        -0.06922   &         0.02236   &         0.34803    \\ 
                        &  $H_{\rm SMS}$  &        0.45657   &        -0.22139   &    {\text{---}}   &         0.23518    \\ 

\hline

  \multirow{2}{*}{ 90}  &  QED            &        0.49792   &        -0.10713   &         0.04057   &         0.43135    \\ 
                        &  $H_{\rm SMS}$  &        0.59422   &        -0.35426   &    {\text{---}}   &         0.23995    \\ 

\hline

  \multirow{2}{*}{ 92}  &  QED            &        0.52397   &        -0.11761   &         0.04570   &         0.45207    \\ 
                        &  $H_{\rm SMS}$  &        0.62927   &        -0.38976   &    {\text{---}}   &         0.23950    \\ 

\hline

  \multirow{2}{*}{ 95}  &  QED            &        0.56762   &        -0.13592   &         0.05472   &         0.48641    \\ 
                        &  $H_{\rm SMS}$  &        0.68812   &        -0.45062   &    {\text{---}}   &         0.23750    \\ 

\hline

  \multirow{2}{*}{100}  &  QED            &        0.65556   &     -0.17564(2)   &         0.07432   &      0.55424(2)    \\ 
                        &  $H_{\rm SMS}$  &        0.80708   &        -0.57769   &    {\text{---}}   &         0.22939    \\

\end{longtable}
}

%% file: table_1_bind_1s1s2p1.tex
{
\renewcommand{\arraystretch}{1.3}
\begin{longtable}{
                  l
                  c
                  S[table-format=-1.5(1)]
                  S[table-format=-1.5(1)]
                  S[table-format=-1.5(1)]
                  S[table-format=-1.5(1)]
                 }
 \caption{\label{tab:1:bind_1s1s2p1} 
         The interelectronic-interaction correction of first order in $1/Z$ to the two-electron part
         of the nuclear recoil contribution to the binding energy of the $1s^22p_{1/2}$ state
         expressed in terms of the dimensionless function $B(\alpha Z)$ defined by Eq.~(\ref{eq:B_alphaZ}). %
         }\\
\hline
\hline
 \multicolumn{1}{c}{\rule{0pt}{1.2em}$Z$}  &  \multicolumn{1}{c}{Approach}  &  \multicolumn{1}{c}{$B_{\rm c}(\alpha Z)$}   &  
             \multicolumn{1}{c}{$B_{\rm tr1}(\alpha Z)$}   &  \multicolumn{1}{c}{$B_{\rm tr2}(\alpha Z)$}   &
             \multicolumn{1}{c}{$B(\alpha Z)$}   \\
\hline
\endfirsthead
\caption{\it (Continued.)}\\
\hline
\hline
 \multicolumn{1}{c}{\rule{0pt}{1.2em}$Z$}  &  \multicolumn{1}{c}{Approach}  &  \multicolumn{1}{c}{$B_{\rm c}(\alpha Z)$}   &  
             \multicolumn{1}{c}{$B_{\rm tr1}(\alpha Z)$}   &  \multicolumn{1}{c}{$B_{\rm tr2}(\alpha Z)$}   &
             \multicolumn{1}{c}{$B(\alpha Z)$}   \\
\hline
\endhead
\hline
\endfoot
\hline
\endlastfoot
                       
  \multirow{2}{*}{  5}  &  QED            &        0.44462   &        -0.00107   &         0.00000   &         0.44355    \\ 
                        &  $H_{\rm SMS}$  &        0.44464   &        -0.00111   &    {\text{---}}   &         0.44353    \\ 

\hline

  \multirow{2}{*}{ 10}  &  QED            &        0.44941   &        -0.00423   &         0.00001   &         0.44519    \\ 
                        &  $H_{\rm SMS}$  &        0.44953   &        -0.00448   &    {\text{---}}   &         0.44505    \\ 

\hline

  \multirow{2}{*}{ 20}  &  QED            &        0.46862   &        -0.01693   &         0.00022   &         0.45191    \\ 
                        &  $H_{\rm SMS}$  &        0.46956   &        -0.01872   &    {\text{---}}   &         0.45084    \\ 

\hline

  \multirow{2}{*}{ 30}  &  QED            &        0.50159   &        -0.03924   &         0.00112   &         0.46348    \\ 
                        &  $H_{\rm SMS}$  &        0.50458   &        -0.04513   &    {\text{---}}   &         0.45946    \\ 

\hline

  \multirow{2}{*}{ 40}  &  QED            &        0.55062   &        -0.07389   &         0.00373   &         0.48046    \\ 
                        &  $H_{\rm SMS}$  &        0.55745   &        -0.08810   &    {\text{---}}   &         0.46936    \\ 

\hline

  \multirow{2}{*}{ 50}  &  QED            &        0.61988   &        -0.12587   &         0.00977   &         0.50378    \\ 
                        &  $H_{\rm SMS}$  &        0.63305   &        -0.15488   &    {\text{---}}   &         0.47816    \\ 

\hline

  \multirow{2}{*}{ 60}  &  QED            &        0.71645   &        -0.20385   &         0.02235   &         0.53496    \\ 
                        &  $H_{\rm SMS}$  &        0.73944   &        -0.25740   &    {\text{---}}   &         0.48204    \\ 

\hline

  \multirow{2}{*}{ 70}  &  QED            &        0.85244   &        -0.32319   &         0.04716   &         0.57640    \\ 
                        &  $H_{\rm SMS}$  &        0.89029   &        -0.41588   &    {\text{---}}   &         0.47441    \\ 

\hline

  \multirow{2}{*}{ 80}  &  QED            &        1.04942   &        -0.51252   &         0.09534   &         0.63224    \\ 
                        &  $H_{\rm SMS}$  &        1.10983   &        -0.66686   &    {\text{---}}   &         0.44297    \\ 

\hline

  \multirow{2}{*}{ 90}  &  QED            &        1.34832   &        -0.82896   &         0.19041   &         0.70977    \\ 
                        &  $H_{\rm SMS}$  &        1.44375   &        -1.08121   &    {\text{---}}   &         0.36255    \\ 

\hline

  \multirow{2}{*}{ 92}  &  QED            &        1.42653   &        -0.91650   &         0.21890   &         0.72893    \\ 
                        &  $H_{\rm SMS}$  &        1.53116   &        -1.19463   &    {\text{---}}   &         0.33653    \\ 

\hline

  \multirow{2}{*}{ 95}  &  QED            &        1.55969   &        -1.06934   &         0.27034   &         0.76069    \\ 
                        &  $H_{\rm SMS}$  &        1.67995   &        -1.39154   &    {\text{---}}   &         0.28841    \\ 

\hline

  \multirow{2}{*}{100}  &  QED            &        1.83558   &     -1.39931(2)   &      0.38725(1)   &      0.82352(2)    \\ 
                        &  $H_{\rm SMS}$  &        1.98799   &        -1.81228   &    {\text{---}}   &         0.17571    \\

\end{longtable}
}

%% file: table_01_bind_1s1s2p1.tex
{
\renewcommand{\arraystretch}{1.3}
\begin{longtable}{
                  l
                  c
                  S[table-format=-3.7]
                  S[table-format=-3.7]
                  S[table-format=-3.6]
                 }
 \caption{\label{tab:01:bind_1s1s2p1} 
         The two-electron part of the nuclear recoil contribution to the binding energy of the $1s^22p_{1/2}$ state.
         The values obtained within the independent electron approximation (to zeroth order in $1/Z$) are given 
         in terms of the dimensionless function $A(\alpha Z)$ defined by Eq.~(\ref{eq:A_alphaZ}). 
         The interelectronic-interaction correction of first order in $1/Z$ is given in terms 
         of the dimensionless function $B(\alpha Z)/Z$ defined by Eq.~(\ref{eq:B_alphaZ}). 
         }\\
\hline
\hline
 \multicolumn{1}{c}{\rule{0pt}{1.2em}$Z$}  &  \multicolumn{1}{c}{Approach}  &  \multicolumn{1}{c}{$A$}   &  
                     \multicolumn{1}{c}{$B/Z$}   &  \multicolumn{1}{c}{$\,\,\,A+B/Z$}   \\
\hline
\endfirsthead
\caption{\it (Continued.)}\\
\hline
\hline
 \multicolumn{1}{c}{\rule{0pt}{1.2em}$Z$}  &  \multicolumn{1}{c}{Approach}  &  \multicolumn{1}{c}{$A$}   &  
                     \multicolumn{1}{c}{$B/Z$}   &  \multicolumn{1}{c}{$\,\,\,A+B/Z$}   \\
\hline
\endhead
\hline
\endfoot
\hline
\endlastfoot
                       
  \multirow{2}{*}{  5}  &  QED            &      -0.077986   &        0.088710   &        0.010723    \\ 
                        &  $H_{\rm SMS}$  &      -0.077986   &        0.088706   &        0.010719    \\ 

\hline

  \multirow{2}{*}{ 10}  &  QED            &      -0.077835   &        0.044519   &       -0.033316    \\ 
                        &  $H_{\rm SMS}$  &      -0.077833   &        0.044505   &       -0.033328    \\ 

\hline

  \multirow{2}{*}{ 20}  &  QED            &      -0.077225   &        0.022595   &       -0.054630    \\ 
                        &  $H_{\rm SMS}$  &      -0.077196   &        0.022542   &       -0.054654    \\ 

\hline

  \multirow{2}{*}{ 30}  &  QED            &      -0.076199   &        0.015449   &       -0.060750    \\ 
                        &  $H_{\rm SMS}$  &      -0.076046   &        0.015315   &       -0.060731    \\ 

\hline

  \multirow{2}{*}{ 40}  &  QED            &      -0.074741   &        0.012011   &       -0.062729    \\ 
                        &  $H_{\rm SMS}$  &      -0.074234   &        0.011734   &       -0.062500    \\ 

\hline

  \multirow{2}{*}{ 50}  &  QED            &      -0.072820   &        0.010076   &       -0.062744    \\ 
                        &  $H_{\rm SMS}$  &      -0.071506   &        0.009563   &       -0.061943    \\ 

\hline

  \multirow{2}{*}{ 60}  &  QED            &      -0.070388   &        0.008916   &       -0.061472    \\ 
                        &  $H_{\rm SMS}$  &      -0.067442   &        0.008034   &       -0.059408    \\ 

\hline

  \multirow{2}{*}{ 70}  &  QED            &      -0.067367   &        0.008234   &       -0.059133    \\ 
                        &  $H_{\rm SMS}$  &      -0.061327   &        0.006777   &       -0.054549    \\ 

\hline

  \multirow{2}{*}{ 80}  &  QED            &      -0.063632   &        0.007903   &       -0.055729    \\ 
                        &  $H_{\rm SMS}$  &      -0.051886   &        0.005537   &       -0.046349    \\ 

\hline

  \multirow{2}{*}{ 90}  &  QED            &      -0.058988   &        0.007886   &       -0.051102    \\ 
                        &  $H_{\rm SMS}$  &      -0.036694   &        0.004028   &       -0.032666    \\ 

\hline

  \multirow{2}{*}{ 92}  &  QED            &      -0.057926   &        0.007923   &       -0.050003    \\ 
                        &  $H_{\rm SMS}$  &      -0.032597   &        0.003658   &       -0.028939    \\ 

\hline

  \multirow{2}{*}{ 95}  &  QED            &      -0.056235   &        0.008007   &       -0.048227    \\ 
                        &  $H_{\rm SMS}$  &      -0.025536   &        0.003036   &       -0.022500    \\ 

\hline

  \multirow{2}{*}{100}  &  QED            &      -0.053123   &        0.008235   &       -0.044887    \\ 
                        &  $H_{\rm SMS}$  &      -0.010645   &        0.001757   &       -0.008888    \\

\end{longtable}
}

%% file: sms_IntEl_v3_resub.bbl
\begin{thebibliography}{10}

\bibitem{Shabaev:1985:588}
V.~M.{~}Shabaev,
\newblock Theor. Math. Phys. {\bf 63},~588 (1985),
\newblock [Teor. Mat. Fiz. {\bf 63}, 394 (1985)].

\bibitem{Shabaev:1988:69}
V.~M.{~}Shabaev,
\newblock Sov. J. Nucl. Phys. {\bf 47},~69 (1988),
\newblock [Yad. Fiz. {\bf 47}, 107 (1988)].

\bibitem{Palmer:1987:5987}
C.~W.~P.{~}Palmer,
\newblock J. Phys. B: At. Mol. Phys. {\bf 20},~5987 (1987).

\bibitem{Tupitsyn:2003:022511}
I.~I.{~}Tupitsyn, V.~M.{~}Shabaev, J.~R.{~}{Crespo L\'opez-Urrutia},
  I.{~}Dragani\'c, R.{~}Soria~Orts, and J.{~}Ullrich,
\newblock Phys. Rev. A {\bf 68},~022511 (2003).

\bibitem{SoriaOrts:2006:103002}
R.{~}Soria~Orts, Z.{~}Harman, J.~R.{~}{Crespo L\'opez-Urrutia},
  A.~N.{~}Artemyev, H.{~}Bruhns, A.~J.~G.{~}Mart\'inez, U.~D.{~}Jentschura,
  C.~H.{~}Keitel, A.{~}Lapierre, V.{~}Mironov, V.~M.{~}Shabaev, H.{~}Tawara,
  I.~I.{~}Tupitsyn, J.{~}Ullrich, and A.~V.{~}Volotka,
\newblock Phys. Rev. Lett. {\bf 97},~103002 (2006).

\bibitem{Korol:2007:022103}
V.~A.{~}Korol and M.~G.{~}Kozlov,
\newblock Phys. Rev. A {\bf 76},~022103 (2007).

\bibitem{Kozhedub:2010:042513}
Y.~S.{~}Kozhedub, A.~V.{~}Volotka, A.~N.{~}Artemyev, D.~A.{~}Glazov,
  G.{~}Plunien, V.~M.{~}Shabaev, I.~I.{~}Tupitsyn, and {\relax
  Th}.{~}St\"ohlker,
\newblock Phys. Rev. A {\bf 81},~042513 (2010).

\bibitem{Gaidamauskas:2011:175003}
E.{~}Gaidamauskas, C.{~}Naz\'e, P.{~}Rynkun, G.{~}Gaigalas, P.{~}J\"onsson, and
  M.{~}Godefroid,
\newblock J. Phys. B: At. Mol. Opt. Phys. {\bf 44},~175003 (2011).

\bibitem{Zubova:2014:062512}
N.~A.{~}Zubova, Y.~S.{~}Kozhedub, V.~M.{~}Shabaev, I.~I.{~}Tupitsyn,
  A.~V.{~}Volotka, G.{~}Plunien, C.{~}Brandau, and {\relax Th}.{~}St\"ohlker,
\newblock Phys. Rev. A {\bf 90},~062512 (2014).

\bibitem{Naze:2014:1197}
C.{~}Naz\'e, S.{~}Verdebout, P.{~}Rynkun, G.{~}Gaigalas, M.{~}Godefroid, and
  P.{~}J\"onsson,
\newblock At. Data Nucl. Data Tables {\bf 100},~1197 (2014).

\bibitem{Zubova:2016:052502}
N.~A.{~}Zubova, A.~V.{~}Malyshev, I.~I.{~}Tupitsyn, V.~M.{~}Shabaev,
  Y.~S.{~}Kozhedub, G.{~}Plunien, C.{~}Brandau, and {\relax Th}.{~}St\"ohlker,
\newblock Phys. Rev. A {\bf 93},~052502 (2016).

\bibitem{Filippin:2017:042502}
L.{~}Filippin, J.{~}Biero\'n, G.{~}Gaigalas, M.{~}Godefroid, and
  P.{~}J\"onsson,
\newblock Phys. Rev. A {\bf 96},~042502 (2017).

\bibitem{Gamrath:2018:38}
S.{~}Gamrath, P.{~}Palmeri, P.{~}Quinet, S.{~}Bouazza, and M.{~}Godefroid,
\newblock J. Quant. Spectrosc. Radiat. Transf. {\bf 218},~38 (2018).

\bibitem{Tupitsyn:2018:022517}
I.~I.{~}Tupitsyn, N.~A.{~}Zubova, V.~M.{~}Shabaev, G.{~}Plunien, and {\relax
  Th}.{~}St\"ohlker,
\newblock Phys. Rev. A {\bf 98},~022517 (2018).

\bibitem{Ekman:2019:433}
J.{~}Ekman, P.{~}J\"onsson, M.{~}Godefroid, C.{~}Naz\'e, G.{~}Gaigalas, and
  J.{~}Biero\'n,
\newblock Comp. Phys. Comm. {\bf 235},~433 (2019).

\bibitem{Shabaev:1998:59}
V.~M.{~}Shabaev,
\newblock Phys. Rev. A {\bf 57},~59 (1998).

\bibitem{Pachucki:1995:1854}
K.{~}Pachucki and H.{~}Grotch,
\newblock Phys. Rev. A {\bf 51},~1854 (1995).

\bibitem{Yelkhovsky:recoil}
A.~S.{~}Yelkhovsky,
\newblock {e-print hep-th/9403095 (1996)}; Zh. Eksp. Fiz. {\bf 110}, 431
  (1996).

\bibitem{Adkins:2007:042508}
G.~S.{~}Adkins, S.{~}Morrison, and J.{~}Sapirstein,
\newblock Phys. Rev. A {\bf 76},~042508 (2007).

\bibitem{Artemyev:1995:1884}
A.~N.{~}Artemyev, V.~M.{~}Shabaev, and V.~A.{~}Yerokhin,
\newblock Phys. Rev. A {\bf 52},~1884 (1995).

\bibitem{Artemyev:1995:5201}
A.~N.{~}Artemyev, V.~M.{~}Shabaev, and V.~A.{~}Yerokhin,
\newblock J. Phys. B: At. Mol. Opt. Phys. {\bf 28},~5201 (1995).

\bibitem{Shabaev:1998:4235}
V.~M.{~}Shabaev, A.~N.{~}Artemyev, T.{~}Beier, G.{~}Plunien, V.~A.{~}Yerokhin,
  and G.{~}Soff,
\newblock Phys. Rev. A {\bf 57},~4235 (1998).

\bibitem{Shabaev:1999:493}
V.~M.{~}Shabaev, A.~N.{~}Artemyev, T.{~}Beier, G.{~}Plunien, V.~A.{~}Yerokhin,
  and G.{~}Soff,
\newblock Phys. Scr. {\bf T80},~493 (1999).

\bibitem{Grotch:1969:350}
H.{~}Grotch and D.~R.{~}Yennie,
\newblock Rev. Mod. Phys. {\bf 41},~350 (1969).

\bibitem{Borie:1982:67}
E.{~}Borie and G.~A.{~}Rinker,
\newblock Rev. Mod. Phys. {\bf 54},~67 (1982).

\bibitem{Aleksandrov:2015:144004}
I.~A.{~}Aleksandrov, A.~A.{~}Shchepetnov, D.~A.{~}Glazov, and V.~M.{~}Shabaev,
\newblock J. Phys. B: At. Mol. Opt. Phys. {\bf 48},~144004 (2015).

\bibitem{Malyshev:2018:085001}
A.~V.{~}Malyshev, R.~V.{~}Popov, V.~M.{~}Shabaev, and N.~A.{~}Zubova,
\newblock J. Phys. B: At. Mol. Opt. Phys. {\bf 51},~085001 (2018).

\bibitem{Malyshev:2017:765}
A.~V.{~}Malyshev, V.~M.{~}Shabaev, D.~A.{~}Glazov, and I.~I.{~}Tupitsyn,
\newblock JETP Lett. {\bf 106},~765 (2017).

\bibitem{Shi:2016:2}
C.{~}Shi, F.{~}Gebert, C.{~}Gorges, S.{~}Kaufmann, W.{~}N\"ortersh\"auser,
  B.~K.{~}Sahoo, A.{~}Surzhykov, V.~A.{~}Yerokhin, J.~C.{~}Berengut, F.{~}Wolf,
  J.~C.{~}Heip, and P.~O.{~}Schmidt,
\newblock Appl. Phys. B {\bf 123},~2 (2016).

\bibitem{Botsi:ICPEAC}
S.{~}Botsi, N.{~}Camus, L.{~}Fechner, T.{~}Pfeifer, and R.{~}Moshammer,
\newblock ICPEAC XXX, Abstract, MO-45 (2017).

\bibitem{Botsi:MS_thesis}
S.{~}Botsi,
\newblock Ms. thesis, University of Heidelberg, 2017.

\bibitem{Furry:1951:115}
W.~H.{~}Furry,
\newblock Phys. Rev. {\bf 81},~115 (1951).

\bibitem{TTGF}
V.~M.{~}Shabaev,
\newblock Phys. Rep. {\bf 356},~119 (2002).

\bibitem{Shabaev:1993:4703}
V.~M.{~}Shabaev,
\newblock J. Phys. B: At. Mol. Opt. Phys. {\bf 26},~4703 (1993).

\bibitem{Shabaev:1994:4489}
V.~M.{~}Shabaev and I.~G.{~}Fokeeva,
\newblock Phys. Rev. A {\bf 49},~4489 (1994).

\bibitem{Yerokhin:2001:032109}
V.~A.{~}Yerokhin, A.~N.{~}Artemyev, V.~M.{~}Shabaev, M.~M.{~}Sysak,
  O.~M.{~}Zherebtsov, and G.{~}Soff,
\newblock Phys. Rev. A {\bf 64},~032109 (2001).

\bibitem{Angeli:2013:69}
I.{~}Angeli and K.~P.{~}Marinova,
\newblock At. Data Nucl. Data Tables {\bf 99},~69 (2013).

\bibitem{Yerokhin:2015:033103}
V.~A.{~}Yerokhin and V.~M.{~}Shabaev,
\newblock J. Phys. Chem. Ref. Data {\bf 44},~033103 (2015).

\bibitem{Johnson:1988:307}
W.~R.{~}Johnson, S.~A.{~}Blundell, and J.{~}Sapirstein,
\newblock Phys. Rev. A {\bf 37},~307 (1988).

\bibitem{Sapirstein:1996:5213}
J.{~}Sapirstein and W.~R.{~}Johnson,
\newblock J. Phys. B: At. Mol. Opt. Phys. {\bf 29},~5213 (1996).

\bibitem{splines:DKB}
V.~M.{~}Shabaev, I.~I.{~}Tupitsyn, V.~A.{~}Yerokhin, G.{~}Plunien, and
  G.{~}Soff,
\newblock Phys. Rev. Lett. {\bf 93},~130405 (2004).

\bibitem{Glazov:2017:46}
D.~A.{~}Glazov, A.~V.{~}Malyshev, A.~V.{~}Volotka, V.~M.{~}Shabaev,
  I.~I.{~}Tupitsyn, and G.{~}Plunien,
\newblock Nucl. Instrum. Methods Phys. Res., Sect. B {\bf 408},~46 (2017).

\bibitem{Shabaev:2017:263001}
V.~M.{~}Shabaev, D.~A.{~}Glazov, A.~V.{~}Malyshev, and I.~I.{~}Tupitsyn,
\newblock Phys. Rev. Lett. {\bf 119},~263001 (2017).

\bibitem{Sturm:2017:4}
S.{~}Sturm, M.{~}Vogel, F.{~}{K\"ohler-Langes}, W.{~}Quint, K.{~}Blaum, and
  G.{~}Werth,
\newblock Atoms {\bf 5},~4 (2017).

\bibitem{Lindenfels:2013:023412}
D.{~}{von Lindenfels}, M.{~}Wiesel, D.~A.{~}Glazov, A.~V.{~}Volotka,
  M.~M.{~}Sokolov, V.~M.{~}Shabaev, G.{~}Plunien, W.{~}Quint, G.{~}Birkl,
  A.{~}Martin, and M.{~}Vogel,
\newblock Phys. Rev. A {\bf 87},~023412 (2013).

\bibitem{Vogel:2019:1800211}
M.{~}Vogel, M.~S.{~}Ebrahimi, Z.{~}Guo, A.{~}Khodaparast, G.{~}Birkl, and
  W.{~}Quint,
\newblock Ann. Phys. (Berlin) {\bf 531},~1800211 (2019).

\bibitem{Yerokhin:2015:233002}
V.~A.{~}Yerokhin and V.~M.{~}Shabaev,
\newblock Phys. Rev. Lett. {\bf 115},~233002 (2015).

\bibitem{Yerokhin:2016:062514}
V.~A.{~}Yerokhin and V.~M.{~}Shabaev,
\newblock Phys. Rev. A {\bf 93},~062514 (2016).

\end{thebibliography}
